\documentclass[11pt,a4paper]{article}
\usepackage{jheppub}
\usepackage{amsmath}
\usepackage{amssymb}
\usepackage{graphicx}
\usepackage{babel}
\usepackage[caption=false]{subfig}
\usepackage{mathtools}
\usepackage{slashed}
\usepackage{mathrsfs} % for mathscr 
\usepackage{xcolor}

\newcommand{\ii}{\text{i}}
\newcommand{\ip}{{\bf 7}}

\definecolor{kblue}{rgb}{0,0.48,0.73}
\definecolor{kred}{rgb}{0.73,0.25,0}
\definecolor{kgreen}{rgb}{0.48,0.73,0}
\definecolor{dgreen}{rgb}{0.2,0.51,0.19}
\definecolor{dred}{rgb}{0.7,0.15,0.09}

\begin{document}
\allowdisplaybreaks[4]

\begin{titlepage}

\begin{center}

{\LARGE \bfseries
The stability of non-Abelian \(A_4\) vortices \\ in \(SO(3)\) gauge theory
}

\vspace{1cm}

\renewcommand{\thefootnote}{\fnsymbol{footnote}}
{%
\hypersetup{linkcolor=black}
Yoshihiko Abe$^{1,2,3}$\footnote[1]{yabe3@keio.jp},
\ 
Tetsutaro Higaki$^{4,5}$\footnote[2]{thigaki@rk.phys.keio.ac.jp},
\ 
Kazuya Murakami$^{4}$\footnote[2]{kazuyamurakami@keio.jp},
\ 
Muneto Nitta$^{5,6,7}$\footnote[2]{mune.nitta@gmail.com},
\ and 
Ryo Yokokura$^{5,6}$\footnote[2]{ryokokur@keio.jp},
}%
\vspace{8mm}

{\itshape%
$^1${Graduate School of Science and Technology, Keio University, Yokohama , Japan}\\
$^2${Keio University Sustainable Quantum Artificial Intelligence Center (KSQAIC), Keio University, Tokyo 108-8345, Japan}\\
$^3${Quantum Computing Center, Keio University, 3-14-1 Hiyoshi, Kohoku-ku, Yokohama, Kanagawa, 223-8522, Japan}\\
$^4${Department of Physics, Keio University, Yokohama 223-8533, Japan}\\
$^5${Reserach and Education Center for Natural Sciences, Keio University,
4-1-1 Hiyoshi, Yokohama, Kanagawa 223-8521, Japan}\\
$^6${Department of Physics,  Keio University,
4-1-1 Hiyoshi, Yokohama, Kanagawa 223-8521, Japan}\\
$^7${International Institute for Sustainability with Knotted Chiral Meta Matter(WPI-SKCM$^2$), Hiroshima University, 1-3-1 Kagamiyama, Higashi-Hiroshima, Hiroshima 739-8531, Japan}
}%

\vspace{8mm}

\end{center}

\begin{abstract}
~
We study the linear energetic stability of non-Abelian vortices in an \(SO(3)\) gauge-Higgs theory with residual gauge group \(A_4\).
In our previous work, we identified two families of such vortices associated with order-two and order-three elements of \(A_4\), and found
Bogomol'nyi--Prasad--Sommerfield-like behavior at special ratios of the Higgs and gauge-boson mass scales.
Here, we investigate their stability against 
\(SO(3)\) gauge and Higgs perturbations by analyzing the second order variation of the vortex tension using analytical and numerical methods. In particular, when all relevant mass scales are equal, we prove linear
stability for all vortices in the order-three family. By contrast, at the same mass point, we prove that a vortex in the
order-two family is unstable. Thus, although both vortex families exhibit Bogomol'nyi--Prasad--Sommerfield-like behavior in their reduced equations at this mass point, their stability properties in the full \(SO(3)\) theory are qualitatively different.

\end{abstract}

\end{titlepage}

\renewcommand{\thefootnote}{\arabic{footnote}}
\setcounter{footnote}{0}
\setcounter{page}{1}

\tableofcontents

\section{Introduction}
Topological solitons or defects are classical field configurations that cannot be continuously deformed into the vacuum while preserving the asymptotic boundary conditions \cite{Manton:2004tk,Weinberg:2012pjx}.
For a defect with a localized core, the fields approach a vacuum manifold \(\mathcal M\) away from the core.
The field values on spheres surrounding or linking the core then define maps from the boundary into \(\mathcal M\), whose homotopy classes characterize the topological sectors \cite{Kibble:1976sj,Mermin:1979zz,Vilenkin:2000jqa,Manton:2004tk,Weinberg:2012pjx}.
In three spatial dimensions, the nontrivial homotopy groups \(\pi_0(\mathcal M)\), \(\pi_1(\mathcal M)\), and \(\pi_2(\mathcal M)\) support  domain walls, vortices or strings, and monopoles, respectively. 
The Kibble mechanism 
\cite{Kibble:1976sj} 
provides one important route to the formation of such defects.
When a system 
undergoes 
a phase transition 
accompanied with spontaneous  symmetry breaking or Higgsing, 
it results in a vacuum manifold.
In such a transition, 
causally disconnected regions 
in space may select different vacuum configurations independently.
When these regions subsequently meet, the fields may be unable to interpolate everywhere within the vacuum manifold, and topological defects may remain.
A closely related nonequilibrium mechanism occurs in rapidly quenched laboratory systems and is known as the Kibble--Zurek mechanism \cite{Zurek:1985qw,Zurek:1996sj}. 
Such Kibble-Zurek mechanism was confirmed in several laboratory experiments \cite{Bowick:1992rz,Hendry:1994ami,Sadler:2006cok,Weiler:2008jpd}. 

Among these topological defects, the present article concerns vortices or cosmic strings. 
Historically, the study of vortices and line defects has played an important role in 
condensed-matter systems, including superfluids, superconductors, liquid crystals, superfluid \({}^3\mathrm{He}\), and spinor Bose--Einstein condensates \cite{Abrikosov:1956sx,Mermin:1979zz,salomaaRMP,Kawaguchi:2012ii}.
In \((1+3)\) dimensions, they extend along one spatial direction and are therefore stringlike.
When such strings are produced during phase transitions in the early universe, they become cosmic strings \cite{Kibble:1976sj,Vilenkin:1982sh,Hindmarsh:1994re,Vilenkin:2000jqa}. 

Vortex strings also arise in relativistic quantum field theories \cite{Nielsen:1973cs,Manton:2004tk,Weinberg:2012pjx}.
In particular, the Abelian-Higgs model provides the canonical example of a finite-tension vortex \cite{Nielsen:1973cs,deVega:1976xbp,Jaffe:1980mj}.
For a finite-energy configuration, the Higgs field approaches the vacuum manifold on a sufficiently large circle, defining a map \(S^1_\infty\to U(1)\) characterized by the winding number
    $n \in \pi_1\bigl[U(1)\bigr] \simeq \mathbb Z$.
This also counts the number of the quantized magnetic fluxes carried by the vortices, 
and it is a topological invariant  fixed by the index theorem \cite{Weinberg:1979er}.

The asymptotic interaction between well-separated vortices depends strongly on the Higgs- and gauge-boson masses \(m_H\) and \(m_G\).
The lighter field has the longer correlation length and dominates the long-range interaction.
The exchange of the Higgs boson yields an attractive interaction, whereas the exchange of the gauge field gives a repulsive interaction \cite{Jacobs:1978ch,Speight:1996px}.
Vortices therefore attract each other for \(m_H<m_G\) and repel each other for \(m_G<m_H\), corresponding to the type-I and type-II regimes, respectively.
At the critical coupling \(m_H=m_G\), 
the force is canceled out.
In such a case, 
the total energy of multiple vortices can be written as a sum of non-negative squares together with a term determined by the topological charge \cite{Bogomolny:1975de,Jaffe:1980mj}.
This decomposition gives the Bogomol'nyi lower bound on the vortex tension.
Configurations that saturate this bound satisfy a set of first-order differential equations, known as the Bogomol'nyi equations.
For positive winding \(n>0\), their tensions satisfy
\begin{align}
    E_n = n E_1,
\end{align}
where \(E_n\) denotes the tension of a winding-\(n\) Bogomol'nyi solution. 
These vorices are called 
Bogomol'nyi-Prasad-Sommerfield (BPS) vortices. 
For straight, static vortices extending along the \(z\)-axis, the equations of motion reduce to equations on the transverse \((x,y)\) plane.
For any prescribed set of \(n\) points in this plane, counted with multiplicity, there exists a static Bogomol'nyi solution whose vortex centers are located at those points \cite{Taubes:1979tm}.
Modulo gauge transformations, these solutions form a \(2n\)-dimensional moduli space parameterized by the vortex centers \cite{Weinberg:1979er,Taubes:1979tm,Samols:1991ne,Stuart:1994yc}.
In supersymmetric field theories, 
BPS vortices preserve a half of supercharges and are called $1/2$ BPS states 
\cite{DiVecchia:1977nxl,Witten:1978mh,Edelstein:1993bb,Hanany:2003hp,Auzzi:2003fs,Eto:2004rz,Eto:2005yh,Eto:2006uw,Eto:2006cx,Tong:2005un,Eto:2006pg,Shifman:2007ce,Shifman:2009zz,Tong:2008qd}.

Corresponding to type-I and II parameter regions, 
Gustafson and Sigal established the linear stability properties of the 
axisymmetric
Abelian-Higgs \(n\)-vortex, in which the \(n\) vortex centers coincide \cite{GustafsonSigal2000}.
The fundamental vortices with \(|n|=1\) are stable for all values of the couplings.
For \(|n|\geq2\), the coincident vortex is stable in the attractive regime and unstable in the repulsive regime.
At critical coupling, no physical perturbation lowers the energy.
The physical zero modes correspond to infinitesimal displacements of the individual vortex centers \cite{GustafsonSigal2000,Stuart:1994yc}.
Thus, topological protection, inter-vortex forces, binding, and local spectral stability are closely related but logically distinct properties.

A qualitatively richer situation arises when the fundamental group $\pi_1({\cal M})$ of the vacuum manifold ${\cal M}$ is non-Abelian 
\cite{Poenaru1977,Mermin:1979zz,Bais:1980vd,Alford:1990mk,Alford:1990ur,Alford:1992yx,Bucher:1991bc,Lo:1993hp,Brekke:1997jj,McGraw:1997nx,Kobayashi:2008pk}.
The vortex charges are then group-valued rather than additive integers.
Their fusion is governed by noncommutative group multiplication, and braiding can transform the vortex charges nontrivially.
When two vortices carrying noncommuting charges intersect, an additional vortex segment may form between them. 
In gauge theories, the group-valued charge is also encoded in the holonomy around the vortex and leads to a non-Abelian generalization of the Aharonov--Bohm effect \cite{Horvathy:1985jr,Wilczek:1989kn,Alford:1992yx}.
We refer to vortices of this type as {\it non-Abelian}  vortices.\footnote{
    The same term is also used for vortices whose topological charge is Abelian but which carry non-Abelian internal orientational moduli as in 
    supersymmetric QCD \cite{Hanany:2003hp,Auzzi:2003fs,Eto:2004rz,Eto:2005yh,Eto:2006cx,Eto:2006uw,Tong:2005un,Eto:2006pg,Shifman:2007ce,Shifman:2009zz,Tong:2008qd} and dense QCD \cite{Balachandran:2005ev,Nakano:2007dr,Nakano:2008dc,Eto:2009kg,Eto:2009bh,Eto:2009tr,Eto:2013hoa}.
    This usage is distinct from the non-Abelian fundamental group and gauge holonomy considered here.
}
Analogous non-Abelian vortex structures also occur in systems with global or order-parameter symmetries, including ordered media and multicomponent condensates such as liquid crystals \cite{Poenaru1977,Mermin:1979zz,vol77,Balachandran:1983pf,Lavrentovich2001,PhysRevX.15.021036,PhysRevE.52.702,L.Radzihovsky_2001,PhysRevE.77.061701}, 
Bose-Einstein condensates \cite{Semenoff:2006vv,Kobayashi:2008pk,Kawaguchi:2012ii,Borgh:2016cco,Mawson:2018klj,Annala:2022bdd,Rajamaki:2023ymv,Kobayashi:2024aip}, 
and neutron star matter~\cite{Masuda:2016vak,Masaki:2021hmk,Masaki:2023rtn,Kobayashi:2022moc,Kobayashi:2022dae,Marmorini:2020zfp,Hattori:2026bvb}.

One concrete gauge-theory realization of such vortices is provided by an
\(SO(3)\) gauge--Higgs system in which the Higgs vacuum preserves \(A_4\),
the rotational symmetry group of a regular tetrahedron, as the residual gauge group~\cite{Abe:2026zrk}.
Higgsing \(SO(3)\) to finite subgroups, including \(A_4\), has been studied in Refs.~\cite{Ovrut:1977cn,Etesi:1996urw,King:2018fke}.
Recently, in Ref.~\cite{Abe:2026zrk}, the present authors constructed finite-tension axisymmetric vortices in a renormalizable \((1+3)\)-dimensional \(SO(3)\) gauge theory with a real spin-three Higgs field and a residual gauge group \(A_4\).
The Higgs vacuum manifold is \(SO(3)/A_4\), whose fundamental group is 
the binary tetrahedral group $\tilde A_4$ 
which is non-Abelian.\footnote{
Global vortices associated with \(A_4\) order and its binary-tetrahedral fundamental group have been studied in  
liquid crystals 
\cite{PhysRevE.52.702,L.Radzihovsky_2001,PhysRevE.77.061701} 
and Bose-Einstein condensates
\cite{Semenoff:2006vv,Kobayashi:2008pk,Mawson:2018klj,Rajamaki:2023ymv,Kobayashi:2024aip}.
}
Two families of axisymmetric vortex solutions were obtained, associated with order-two and order-three elements of the residual gauge group.
To describe their asymptotic structure, let us first fix a vacuum \(\phi_{\mathrm{vac}}\). Let \(r_p\in A_4\) be an element of order \(p\) for each \(p=2,3\), and choose \(Y_p\in\mathfrak{so}(3)\) such that
\begin{align}
    \exp(2\pi Y_p)=r_p.
\end{align}
Let \(\rho\) denote the spin-three representation of $SO(3)$.
If the internal rotation generated by \(Y_p\) winds \(q\) times around the vortex as \(\theta\) varies from \(0\) to \(2\pi\), the asymptotic Higgs field takes the form
\begin{align}
    \phi_{p,q}(\theta)
    := \rho\bigl(\exp(q\theta Y_p)\bigr)\phi_{\mathrm{vac}}.
\end{align}
Since \(\exp(2\pi qY_p) = r_p^q \in A_4\), this defines a closed loop in \(SO(3)/A_4\), and the corresponding asymptotic gauge holonomy is \(r_p^q\).
For \(p=2\), we write \(q=N\) and call the resulting solutions the order-two vortex family; for \(p=3\), we write \(q=M\) and call them the order-three vortex family.
The integers \(N\) and \(M\) therefore count repeated internal rotations, in direct analogy with the Abelian asymptotic field \(e^{\ii n\theta}\phi_{\mathrm{vac}}\).
Unlike the Abelian winding number, however, \(N\) and \(M\) are integer labels of the axisymmetric ans\"atze.
Because the topological charge takes values in a finite non-Abelian group, different values of these integers need not represent distinct topological sectors \cite{Abe:2026zrk}.
As in the Abelian-Higgs model, the tensions of these vortices depend on the relative Higgs- and gauge-boson mass scales.
At particular mass ratios, the reduced energy functional and field equations governing each of the two axisymmetric vortex families coincide with those of the critically coupled Abelian-Higgs model \cite{Abe:2026zrk}.
Consequently, the radial profiles obey the same first-order equations that characterize critically coupled Abelian-Higgs vortices, and their tensions can be evaluated exactly.
This correspondence further allows the critically coupled Abelian-Higgs multi-vortex solutions, with arbitrarily prescribed centers in the transverse plane, to be used to construct solutions of the full \(SO(3)\) field equations.
In the Abelian-Higgs model, every such multi-vortex solution saturates
the Bogomol'nyi lower bound and therefore minimizes the energy within its fixed winding sector.
For the \(A_4\) vortices obtained from this construction, however, the reduction to the critically coupled Abelian-Higgs model applies only within the restricted classes of fields used to construct the two vortex families.
It therefore does not show that the energy cannot be lowered by general \(SO(3)\) gauge and Higgs perturbations.
Determining whether such perturbations contain negative modes is the central problem addressed in this article.

We study this question using the Hessian of the vortex tension.
Let \(\Psi_\star\) be a static vortex configuration and \(\delta\Psi\) a perturbation.
Since \(\Psi_\star\) is stationary,
\begin{align}
    E[\Psi_\star+s\delta\Psi]
    = E[\Psi_\star] + \frac{s^2}{2} \left\langle\delta\Psi,L_{\Psi_\star}\delta\Psi\right\rangle + \mathcal O(s^3),
\end{align}
where \(L_{\Psi_\star}\) is the Hessian operator governing the quadratic
fluctuations about \(\Psi_\star\), and \(\langle\cdot,\cdot\rangle\)
denotes the corresponding inner product, including integration over the
plane transverse to the vortex.
If \(L_{\Psi_\star}\) has a negative eigenvalue, a perturbation along the corresponding eigenmode lowers the tension at quadratic order and signals linear energetic instability.
Throughout this paper, linear energetic stability means the absence of physical negative modes of the static Hessian.
Following the strategy of Ref.~\cite{GustafsonSigal2000}, we first fix the gauge and expand the perturbations in Fourier modes of the polar angle.
The two-dimensional fluctuation problem then reduces to one-dimensional radial problems.
We study these radial problems using both analytical and numerical methods.

The remainder of this article is organized as follows.
In Sec.~\ref{sec:model}, we introduce the \(SO(3)\) gauge--Higgs model and formulate the stability problem using the gauge-fixed Hessian.
In Secs.~\ref{sec:order-three-stability} and~\ref{sec:order-two-stability}, we study the stability of the order-three and order-two vortex families, respectively, using both analytical and numerical methods.
We summarize and discuss the results in Sec.~\ref{sec:summary}.

\section{Model and linearized fluctuation problem}
\label{sec:model}
In this section, we introduce the model and rewrite the static tension in terms of dimensionless variables.
We then derive the Hessian governing linear stability about a static vortex background and, following Gustafson and Sigal~\cite{GustafsonSigal2000}, eliminate its infinite-dimensional gauge kernel by imposing the background gauge condition.

\subsection{Action and dimensionless parameters}
We consider an \(SO(3)\) gauge theory in \((1+3)\) dimensions coupled to a real Higgs field in the seven-dimensional, i.e., real spin-three irreducible representation denoted by $\ip$. 
(See also Ref.~\cite{Abe:2026zrk} for details.)

A convenient realization of the representation space is the space of real homogeneous harmonic polynomials of degree three,
\begin{align}
    \mathcal H_3 := \left\{f\in\mathbb R[\mathcal X,\mathcal Y,\mathcal Z] \,\middle|\, \deg f=3,\quad \Delta f=0\right\}.
\end{align}
The \(SO(3)\) action is induced by rotations of the arguments,
\begin{align}
    \bigl(\rho(g)f\bigr)(\boldsymbol{\mathcal X}) := f(g^{-1}\boldsymbol{\mathcal X}), \qquad g\in SO(3),
\end{align}
where \(\boldsymbol{\mathcal X}:=(\mathcal X,\mathcal Y,\mathcal Z)^{\mathsf T}\).
We equip \(\mathcal H_3\) with the \(SO(3)\)-invariant inner product
in the $\ip$ representation,
\begin{align}
    \langle f_1,f_2\rangle_\ip := \frac{1}{4\pi}\int_{\mathcal S^2}f_1(\boldsymbol{\mathcal X})f_2(\boldsymbol{\mathcal X})\,d\Omega,
    \label{eq:H3-inner-product}
\end{align}
where ${\mathcal S^2}$ is a two-dimensional sphere in the representation space,
\begin{align}
    \mathcal S^2 := \left\{(\mathcal X,\mathcal Y,\mathcal Z)\in\mathbb R^3 \,\middle|\, \mathcal X^2+\mathcal Y^2+\mathcal Z^2=1\right\},
\end{align}
and \(d\Omega\) is its standard area element.

The action of the theory is
\begin{align}
    S = \int d^4x\left[-\frac{1}{4e^2}G_{\mu\nu}^aG^{a\mu\nu}-\frac{1}{2}\langle D_\mu\phi,D^\mu\phi\rangle_\ip-V(\phi)\right].
\end{align}
Here \(e>0\) is the gauge coupling, \(G_{\mu\nu}\) is the field-strength tensor of the \(SO(3)\) gauge field, \(\phi\) is a real spin-three Higgs field taking values in \(\mathcal H_3\), and \(V(\phi)\) is its potential.
We choose a basis \(\mathrm{i}T_a\) (\(a=1,2,3\)) of the Lie algebra \(\mathfrak{so}(3)\) in its defining three-dimensional representation as
\begin{align}
    \ii T_1 &=
    \begin{pmatrix}
        0 & 0 & 0\\
        0 & 0 & 1\\
        0 & -1 & 0
    \end{pmatrix},
    &
    \ii T_2 &=
    \begin{pmatrix}
        0 & 0 & -1\\
        0 & 0 & 0\\
        1 & 0 & 0
    \end{pmatrix},
    &
    \ii T_3 &=
    \begin{pmatrix}
        0 & 1 & 0\\
        -1 & 0 & 0\\
        0 & 0 & 0
    \end{pmatrix}.
    \label{eq:so3-generators}
\end{align}
With this convention, the Hermitian generators \(T_a\) satisfy
\begin{align}
    [T_a,T_b] = \ii\varepsilon^{abc}T_c,
\end{align}
Here, \(\varepsilon^{abc}\) is the totally antisymmetric tensor with \(\varepsilon^{123}=1\).
Equivalently, the real antisymmetric generators \(\ii T_a\) obey
\begin{align}
    [\ii T_a,\ii T_b] = -\varepsilon^{abc}\ii T_c.
\end{align}
Writing the gauge field and its field strength as
\begin{align}
    W_\mu = W_\mu^aT_a, \qquad G_{\mu\nu} = G_{\mu\nu}^aT_a,
\end{align}
we define
\begin{align}
    G_{\mu\nu} := \partial_\mu W_\nu - \partial_\nu W_\mu + \ii[W_\mu,W_\nu].
\end{align}
In components, this becomes
\begin{align}
    G_{\mu\nu}^a = \partial_\mu W_\nu^a - \partial_\nu W_\mu^a - \varepsilon^{abc}W_\mu^bW_\nu^c.
\end{align}

If we denote the spin-three representation matrices of the infinitesimal \(SO(3)\) transformations by \(X_a\), the covariant derivative of the Higgs field can be written as
\begin{align}
    D_\mu\phi := \partial_\mu\phi + W_\mu^aX_a\phi.
\end{align}

The Higgs potential is constructed from the independent quadratic and quartic \(SO(3)\)-invariant polynomials on \(\mathcal H_3\).
We define the non-negative quartic \(SO(3)\)-invariant polynomial by
\begin{align}
    Q(\phi) := \mathcal K^{-1}I_4(\phi) - \langle\phi,\phi\rangle_\ip^2, 
\end{align}
where
\begin{align}
    I_4(\phi) &:= \frac{1}{4\pi}\int_{\mathcal S^2}\phi(\boldsymbol{\mathcal X})^4\,d\Omega, \\
    \mathcal K &:= \min_{\phi\in\mathcal H_3\setminus\{0\}}\frac{I_4(\phi)}{\langle\phi,\phi\rangle_\ip^2}
    =\frac{315}{143}.
\end{align}
We take the Higgs potential to be
\begin{align}
    V(\phi) := 
\gamma Q(\phi) -\mu^2 \langle\phi, \phi\rangle_\ip + \lambda 
\langle\phi, \phi\rangle_\ip^2
+ \frac{\mu^4}{4\lambda}=
\gamma Q(\phi)+\lambda\bigl(\langle\phi, \phi\rangle_\ip - v^2\bigr)^2, 
\quad v^2:=\frac{\mu^2}{2\lambda},
\end{align}
where \(\mu^2>0\), \(\gamma>0\), and \(\lambda>0\).
The vacuum conditions are therefore
\begin{align}
    \langle\phi,\phi\rangle_\ip=v^2, \qquad Q(\phi)=0.
\end{align}
Up to an \(SO(3)\) rotation, a vacuum configuration is proportional to \(\mathcal X\mathcal Y\mathcal Z\), with its normalization fixed by \(\langle\phi, \phi\rangle_\ip=v^2\).
The un-Higgsed subgroup is therefore \(A_4\subset SO(3)\).

Under a local gauge transformation \(g(x)\in SO(3)\), the Higgs and gauge fields transform as
\begin{align}
    \phi &\longmapsto \rho(g)\phi, \\
    W_\mu &\longmapsto g W_\mu g^{-1} + \ii(\partial_\mu g)g^{-1}.
\end{align}
For an infinitesimal gauge transformation
\begin{align}
    g(x) = \exp\!\left[\alpha^a(x)\ii T_a\right]
    = 1 + \alpha^a(x)\ii T_a + \mathcal O(\alpha^2),
\end{align}
gauge field and Higgs field transform as
\begin{align}
    \phi &\longmapsto \phi + \alpha^a X_a\phi,\\
    W_\mu^a &\longmapsto W_\mu^a - \partial_\mu\alpha^a+\varepsilon^{abc}W_\mu^b\alpha^c.
    \label{eq:infinitesimal-gauge-transformation}
\end{align}

In this article, we study the linear stability of straight vortices extending along the \(z\)-axis.
We therefore consider static configurations that are invariant under translations along \(z\), and choose
\begin{align}
    \partial_t W_\mu = \partial_z W_\mu = 0, \qquad
    \partial_t \phi = \partial_z \phi = 0, \qquad
    W_0 = W_z = 0.
\end{align}
Under these assumptions, the energy per unit length along the \(z\)-axis, namely the vortex tension, is
\begin{align}
    E = \int_{\mathbb R^2}dx dy
    \left[
        \frac{1}{4e^2}G_{ij}^aG_{ij}^a + \frac{1}{2}\langle D_i\phi,D_i\phi\rangle_\ip + V(\phi)
    \right],
    \qquad i,j=1,2.
\end{align}

To express the tension in dimensionless form, we introduce the characteristic mass scales
\begin{align}
    m_+ := 2\mu, \qquad
    m_- := \frac{4}{3\sqrt5}\frac{\mu\sqrt{\gamma}}{\sqrt{\lambda}},
    \qquad
    m_G := \frac{\sqrt2\,\mu e}{\sqrt{\lambda}}=2ev,
\end{align}
Here \(m_+\) and \(m_-\) are the masses of the two Higgs modes, while \(m_G\) is the gauge-boson mass \cite{Abe:2026zrk}.
We then define the two dimensionless parameters
\begin{align}
    \kappa := \frac{m_+}{m_G}=\frac{\sqrt{2\lambda}}{e}, \qquad
    \eta := \frac{m_+}{m_-}
    = \frac{3\sqrt5}{2}\sqrt{\frac{\lambda}{\gamma}}.
\end{align}

We use dimensionless coordinates and fields defined by
\begin{align}
    \widetilde{x}^{\,i}:=evx^i, \qquad
    \widetilde{\phi}:=\frac{\phi}{v}, \qquad
    \widetilde{W}_i^a:=\frac{W_i^a}{ev}, \qquad i=1,2.
\end{align}
We denote differentiation with respect to the dimensionless coordinates by
\begin{align}
    \widetilde{\partial_i}:=\frac{\partial}{\partial\widetilde{x}^{\,i}}.
\end{align}
The corresponding dimensionless field strength and covariant derivative are
\begin{align}
    \widetilde{G}_{ij}^a
    &:=
    \widetilde{\partial_i}\widetilde{W}_j^a
    -\widetilde{\partial_j}\widetilde{W}_i^a
    -\varepsilon^{abc}\widetilde{W}_i^b\widetilde{W}_j^c, \\
    \widetilde{D_i}\widetilde{\phi}
    &:=
    \widetilde{\partial_i}\widetilde{\phi}
    +\widetilde{W}_i^aX_a\widetilde{\phi}.
\end{align}
The dimensional quantities are related to the dimensionless ones by
\begin{align}
    G_{ij}^a = (ev)^2\widetilde{G}_{ij}^a, \qquad D_i\phi = ev^2\widetilde{D}_i\widetilde{\phi}, \qquad d^2x = \frac{d^2\widetilde{x}}{e^2v^2}.
\end{align}
The vortex tension can therefore be written as
\begin{align}
    E = v^2\widetilde{E},
\end{align}
where
\begin{align}
    \widetilde{E}
    := \int_{\mathbb R^2}d^2\widetilde{x}
    \left[
        \frac14\widetilde{G}_{ij}^a\widetilde{G}_{ij}^a + \frac12\langle\widetilde{D}_i\widetilde{\phi},\widetilde{D}_i\widetilde{\phi}\rangle_\ip + \widetilde{V}(\widetilde{\phi})
    \right],
\end{align}
and the dimensionless Higgs potential is
\begin{align}
    \widetilde{V}(\widetilde{\phi})
    := \frac{45\kappa^2}{8\eta^2}Q(\widetilde{\phi}) + \frac{\kappa^2}{2}\left(\langle\widetilde{\phi}, \widetilde{\phi}\rangle_\ip - 1\right)^2.
    \label{eq:dimensionless-potential}
\end{align}
Since the overall factor \(v^2\) is positive, it does not affect the classical field equations or the signs of the Hessian eigenvalues.
In the remainder of this article, we omit the tildes and use dimensionless quantities unless otherwise stated.

For an adjoint-valued field \(\alpha = \alpha^aT_a\), we define the covariant derivative in the adjoint representation by
\begin{align}
    \bigl(\mathcal D_i\alpha\bigr)^a &:= \partial_i \alpha^a - \varepsilon^{abc} W_i^b \alpha^c.
    \label{eq:adjoint-covariant-derivative}
\end{align}
In terms of the dimensionless quantities introduced above, the Euler--Lagrange equations take the form
\begin{align}
    0 &= -\bigl(\mathcal D_i G_{ij}\bigr)^a + \langle X_a \phi,D_j\phi\rangle_\ip, \label{eq:dimensionless-gauge-equation}\\
    0 &= -D_i D_i \phi + \frac{45\kappa^2}{8\eta^2}\nabla Q(\phi) + 2\kappa^2 \bigl(\langle\phi, \phi\rangle_\ip - 1\bigr)\phi. \label{eq:dimensionless-Higgs-equation}
\end{align}
Here, \(\nabla Q(\phi) \in \mathcal{H}_3\) denotes the gradient of the function \(Q:\mathcal H_3 \to \mathbb{R}\) in the representation space with respect to the inner product \(\langle\cdot, \cdot\rangle_\ip\) on \(\mathcal H_3\), namely,
\begin{align}
    \left.\frac{d}{ds}\right|_{s=0} Q(\phi + s \xi) = \langle \nabla Q(\phi), \xi \rangle_\ip \qquad
    \text{for all } \xi \in \mathcal{H}_3.
\end{align}

\subsection{Hessian and gauge fixing}
\label{subsec:hessian-gauge-fixing}
We first define the Hessian abstractly as the second order variation of the energy. We then collect the ingredients needed to evaluate it explicitly and finally derive its gauge- and Higgs-field components.

Let \((W,\phi)\) be a static solution of the Euler--Lagrange equations \eqref{eq:dimensionless-gauge-equation} and \eqref{eq:dimensionless-Higgs-equation}.
We introduce real fluctuations of the gauge and Higgs fields by
\begin{align}
    W_i^a \longmapsto W_i^a + sB_i^a,
    \qquad
    \phi \longmapsto \phi + s\xi.
\end{align}
Here, \(s\) is a dimensionless infinitesimal parameter.
For two fluctuations \((B,\xi)\) and \((C,\zeta)\), we use the following inner product
\begin{align}
    \bigl\langle (C,\zeta),(B,\xi)\bigr\rangle
    := \int_{\mathbb R^2}d^2x \left[C_i^a B_i^a + \langle\zeta, \xi\rangle_\ip\right].
\end{align}
The Hessian \(L\) about the background \((W,\phi)\) is defined by
\begin{align}
    &\left.\frac{\partial^2}{\partial s\,\partial t}\right|_{s=t=0} E\bigl(W+sB+tC,\phi+s\xi+t\zeta\bigr) \notag\\
    &\qquad\qquad = \bigl\langle (C,\zeta), L(B,\xi)\bigr\rangle \notag\\
    &\qquad\qquad = \int_{\mathbb R^2}d^2x \left[C_j^a\bigl(L_W(B,\xi)\bigr)_j^a + \bigl\langle\zeta, L_\phi(B,\xi)\bigr\rangle_\ip\right].
\end{align}
To derive the explicit form of \(L\), we first expand the field strength
and the covariant derivative to first order in the fluctuations:
\begin{align}
    G_{ij}^a(W + sB) &= G_{ij}^a + s\left[\bigl(\mathcal D_iB_j\bigr)^a - \bigl(\mathcal D_jB_i\bigr)^a\right] + \mathcal O(s^2), \notag\\
    D_i^{W + sB}(\phi + s\xi) &= D_i\phi + s\left(D_i\xi + B_i^a X_a\phi\right) + \mathcal O(s^2).
\end{align}
Here \(\mathcal D_i\) is the adjoint covariant derivative defined in \eqref{eq:adjoint-covariant-derivative}, evaluated on the background gauge field \(W\).
Assuming that the fluctuations decay sufficiently rapidly at spatial infinity so that the boundary terms vanish, we expand \(E\) to second order and integrate by parts to obtain
\begin{align}
    \bigl(L_W(B,\xi)\bigr)_j^c
    &= \bigl(\mathcal D_i^\dagger(\mathcal D_i B_j - \mathcal D_j B_i)\bigr)^c -  \varepsilon^{abc}G_{ij}^a B_i^b \notag \\
    &\qquad + \left\langle X_c\phi, D_j\xi + B_j^b X_b\phi\right\rangle_\ip + \left\langle D_j\phi, X_c\xi \right\rangle_\ip,\\
    L_\phi(B,\xi)
    &= - D_i\left(D_i\xi + B_i^a X_a\phi\right) -  B_i^a X_a D_i\phi + H_V(\phi)\xi.
\end{align}
Here \(\dagger\) denotes the formal adjoint, and \(H_V(\phi)\) denotes the Hessian of the Higgs potential.
For the adjoint covariant derivative,
\begin{align}
    \mathcal D_i^\dagger = -\mathcal D_i.
\end{align}
Indeed, for adjoint-valued fields \(u\) and \(v\),
\begin{align}
    \int_{\mathbb R^2}d^2x\, u^a\bigl(\mathcal D_i v\bigr)^a
    &= -\int_{\mathbb R^2}d^2x\,\bigl(\mathcal D_i u\bigr)^a v^a \notag\\
    &= \int_{\mathbb R^2}d^2x\,\bigl(\mathcal D_i^\dagger u\bigr)^a v^a.
\end{align}
The Hessian \(H_V(\phi)\) is defined by
\begin{align}
    H_V(\phi)\xi := \left.\frac{d}{ds}\right|_{s=0} \nabla V(\phi+s\xi).
\end{align}
For the dimensionless potential \eqref{eq:dimensionless-potential}, this gives
\begin{align}
    H_V(\phi)\xi &= \frac{45\kappa^2}{8\eta^2} H_{Q}(\phi)\xi
    + 2\kappa^2\bigl(\langle\phi, \phi\rangle_\ip - 1\bigr)\xi + 4\kappa^2 \langle\phi, \xi\rangle_\ip\phi,
\end{align}
where
\begin{align}
    H_Q(\phi)\xi := \left.\frac{d}{ds}\right|_{s=0}\nabla Q(\phi + s\xi).
\end{align}

It is noted that the Hessian has an infinite-dimensional kernel generated by infinitesimal gauge transformations.
As shown in \eqref{eq:infinitesimal-gauge-transformation}, an infinitesimal gauge transformation with parameter \(s\alpha^a(x)\) changes the background fields as
\begin{align}
    W_i^a &\longmapsto W_i^a - s\bigl(\mathcal D_i\alpha\bigr)^a + \mathcal O(s^2), \\
    \phi &\longmapsto \phi + s\alpha^a X_a\phi + \mathcal O(s^2).
\end{align}
Infinitesimal gauge transformations do not change the physical state and hence leave the tension invariant. Therefore, the corresponding fluctuations
\begin{align}
    R\alpha :=
    \begin{pmatrix}
        - \mathcal D_i\alpha\\[1mm]
        \alpha^aX_a\phi
    \end{pmatrix}
\end{align}
are zero modes of the Hessian:
\begin{align}
    L(R\alpha) = 0.
\end{align}

To remove these unphysical gauge zero modes, we restrict the fluctuation space to directions orthogonal to the infinitesimal gauge orbit,
\begin{align}
    \bigl\langle R\alpha, (B,\xi)\bigr\rangle = 0 \qquad
    \text{for every }\alpha.
\end{align}
Using \(\mathcal D_i^\dagger=-\mathcal D_i\), this inner product becomes
\begin{align}
    \bigl\langle R\alpha,(B,\xi)\bigr\rangle
    = -\int_{\mathbb R^2}d^2x\, \alpha^a \left[\bigl(\mathcal D_i^\dagger B_i\bigr)^a - \langle X_a\phi, \xi\rangle_\ip\right].
\end{align}
Thus, this orthogonality condition is equivalent to the background gauge condition
\begin{align}
    \Gamma^a(B,\xi) := \bigl(\mathcal D_i^\dagger B_i\bigr)^a - \langle X_a\phi, \xi\rangle_\ip = 0.
    \label{eq:background-gauge-condition}
\end{align}

Following Ref.~\cite{GustafsonSigal2000}, we define the gauge-fixed Hessian \(\widetilde L\) by adding the corresponding non-negative term:
\begin{align}
    \bigl\langle (C,\zeta), \widetilde L(B,\xi)\bigr\rangle
    := \bigl\langle(C,\zeta), L(B,\xi)\bigr\rangle
    + \int_{\mathbb R^2}d^2x\,\Gamma^a(C,\zeta)\Gamma^a(B,\xi).
\end{align}
The additional term vanishes for fluctuations satisfying \eqref{eq:background-gauge-condition} and therefore does not change the Hessian on this subspace.
Since the gauge parameter is an arbitrary function of position, the gauge zero modes form an infinite-dimensional subspace.
This makes it difficult, especially in numerical calculations, to distinguish physical zero modes, such as translational modes, from the gauge zero modes.
The gauge-fixed Hessian removes this difficulty by lifting the gauge zero modes away from zero without affecting the physical zero modes or negative modes.

A direct calculation gives
\begin{align}
    \begin{aligned}
        \bigl(\widetilde L_W(B, \xi)\bigr)_j^c
        &= -\bigl(\mathcal D_i\mathcal D_i B_j\bigr)^c
        - 2\varepsilon^{abc}G_{ij}^aB_i^b
        + \langle X_c\phi, X_b\phi\rangle_\ip B_j^b
        - 2\langle X_c D_j\phi, \xi\rangle_\ip, \\[1mm]
        \widetilde L_\phi(B,\xi)
        &= - D_i D_i\xi
        - 2B_i^a X_a D_i\phi
        + H_V(\phi)\xi
        + \langle X_a\phi, \xi\rangle_\ip X_a\phi.
    \end{aligned}
\end{align}
In particular, the gauge-fixed Hessian contains no mixed first derivatives of the fluctuations.
After the angular Fourier decomposition, the fluctuation problem therefore reduces to a set of radial Schrödinger-type eigenvalue problems.

\section{Stability of the order-three vortex family}
\label{sec:order-three-stability}
In this section, we analyze the linear stability of the order-three vortex family constructed in Ref.~\cite{Abe:2026zrk}.
After introducing the vortex ansatz and the Fourier decomposition of the gauge-fixed Hessian, we study its spectrum numerically and prove linear stability at \(\kappa=\eta=1\) for arbitrary \(M\).

\subsection{Ansatz and complex basis}
\label{subsec:order-three-ansatz-complex-basis}
We first review the order-three vortex ansatz and introduce a complex basis convenient for the subsequent fluctuation analysis.

We choose \(r_3\) to be the rotation by \(2\pi/3\) about the direction \((1,1,1)/\sqrt{3}\).
To describe this rotation on the gauge and Higgs fields, it is convenient to introduce
\begin{align}
    T_\Delta &:= \frac{T_1+T_2+T_3}{\sqrt{3}}, &
    X_\Delta &:= \frac{X_1+X_2+X_3}{\sqrt{3}}.
\end{align}
With these conventions, the order-three element \(r_3\) is explicitly
\begin{align}
    r_3 &:= \exp\left(\frac{2\pi\ii}{3}T_\Delta\right)
    =
    \begin{pmatrix}
        0 & 1 & 0\\
        0 & 0 & 1\\
        1 & 0 & 0
    \end{pmatrix}, &
    r_3^3 &= \boldsymbol{1}_3.
\end{align}

Since \(\mathcal H_3\) carries the spin-three representation of \(SO(3)\), its restriction to rotations about the \((1,1,1)/\sqrt{3}\) axis decomposes over \(\mathbb R\) into one one-dimensional invariant subspace and three two-dimensional invariant subspaces, corresponding to angular-momentum quantum numbers \(q=0,1,2,3\).
Equivalently, one can choose a real orthonormal basis \(\{f_1,\ldots,f_7\}\) of \(\mathcal H_3\) such that
\begin{align}
    X_\Delta f_1 &= 0, \notag\\
    X_\Delta f_{2q} &= -q f_{2q+1}, &
    X_\Delta f_{2q+1} &= q f_{2q},
    \qquad q=1,2,3.
    \label{eq:order-three-XDelta-real-basis}
\end{align}
Thus, on each two-dimensional subspace \(\operatorname{span}_{\mathbb R}\{f_{2q},f_{2q+1}\}\), \(X_\Delta\) acts as the generator of a rotation with weight \(q\).

One explicit orthonormal basis with this property is
\begin{align}
    f_1 &:= \frac{\sqrt{21}}{9}\Bigl(
        - 2\mathcal X^3
        + 3\mathcal X^2\mathcal Y
        + 3\mathcal X^2\mathcal Z
        + 3\mathcal X\mathcal Y^2
        + 15\mathcal X\mathcal Y\mathcal Z
        + 3\mathcal X\mathcal Z^2
    \nonumber \\ &\hspace{27mm}
        - 2\mathcal Y^3
        + 3\mathcal Y^2\mathcal Z
        + 3\mathcal Y\mathcal Z^2
        - 2\mathcal Z^3
    \Bigr), \\[1mm]
    f_2 &:= \frac{\sqrt{7}}{6}\Bigl(
        - \mathcal X^3
        - 6\mathcal X^2\mathcal Y
        - 3\mathcal X^2\mathcal Z
        - 6\mathcal X\mathcal Y^2
        + 9\mathcal X\mathcal Z^2
        - \mathcal Y^3
        - 3\mathcal Y^2\mathcal Z
        + 9\mathcal Y\mathcal Z^2
        + 2\mathcal Z^3
    \Bigr), \\[1mm]
    f_3 &:= \frac{\sqrt{21}}{6}\Bigl(
        \mathcal X^3
        + 4\mathcal X^2\mathcal Y
        + 5\mathcal X^2\mathcal Z
        - 4\mathcal X\mathcal Y^2
        + \mathcal X\mathcal Z^2
        - \mathcal Y^3
        - 5\mathcal Y^2\mathcal Z
        - \mathcal Y\mathcal Z^2
    \Bigr), \\[1mm]
    f_4 &:= \frac{\sqrt{35}}{6} \Bigl(
        \mathcal X^3
        - 3\mathcal X^2\mathcal Y
        + 3\mathcal X^2\mathcal Z
        - 3\mathcal X\mathcal Y^2
        + \mathcal Y^3
        + 3\mathcal Y^2\mathcal Z
        - 2\mathcal Z^3
    \Bigr), \\[1mm]
    f_5 &:= \frac{\sqrt{105}}{6} \Bigl(
        \mathcal X^3
        + \mathcal X^2\mathcal Y
        - \mathcal X^2\mathcal Z
        - \mathcal X\mathcal Y^2
        - 2\mathcal X\mathcal Z^2
        - \mathcal Y^3
        + \mathcal Y^2\mathcal Z
        + 2\mathcal Y\mathcal Z^2
    \Bigr), \\[1mm]
    f_6 &:= \frac{\sqrt{105}}{18} \Bigl(
        2\mathcal X^3
        - 3\mathcal X^2\mathcal Y
        - 3\mathcal X^2\mathcal Z
        - 3\mathcal X\mathcal Y^2
        + 12\mathcal X\mathcal Y\mathcal Z
        - 3\mathcal X\mathcal Z^2
        \nonumber 
    \\ &\hspace{27mm}
        + 2\mathcal Y^3
        - 3\mathcal Y^2\mathcal Z
        - 3\mathcal Y\mathcal Z^2
        + 2\mathcal Z^3
    \Bigr), \\[1mm]
    f_7 &:= \frac{\sqrt{35}}{2} \Bigl(
        \mathcal X^2\mathcal Y
        - \mathcal X^2\mathcal Z
        - \mathcal X\mathcal Y^2
        + \mathcal X\mathcal Z^2
        + \mathcal Y^2\mathcal Z
        - \mathcal Y\mathcal Z^2
    \Bigr).
\end{align}
These polynomials are orthonormal with respect to \eqref{eq:H3-inner-product}.
The action of the infinitesimal generators on \(\mathcal H_3\) can also be written directly as differential operators:
\begin{align}
    X_1 &= \mathcal Y\frac{\partial}{\partial\mathcal Z} - \mathcal Z\frac{\partial}{\partial\mathcal Y}, \notag\\
    X_2 &= \mathcal Z\frac{\partial}{\partial\mathcal X} - \mathcal X\frac{\partial}{\partial\mathcal Z}, \notag\\
    X_3 &= \mathcal X\frac{\partial}{\partial\mathcal Y} - \mathcal Y\frac{\partial}{\partial\mathcal X}.
\end{align}
Hence
\begin{align}
    X_\Delta = \frac{1}{\sqrt{3}}
    \left[
        (\mathcal Z-\mathcal Y)\frac{\partial}{\partial\mathcal X}
        +(\mathcal X-\mathcal Z)\frac{\partial}{\partial\mathcal Y}
        +(\mathcal Y-\mathcal X)\frac{\partial}{\partial\mathcal Z}
    \right].
\end{align}
Applying this operator to the polynomials above directly verifies \eqref{eq:order-three-XDelta-real-basis}.

The order-three vortex ansatz constructed in Ref.~\cite{Abe:2026zrk}, with \(M\in\mathbb Z_{>0}\), is
\begin{align}
    W &= -\frac{M}{3}K(r)T_\Delta\,d\theta, \label{eq:order-three-gauge-ansatz} \\
    \phi &= \frac{\sqrt{5}}{3}b(r)f_1 + \frac{2}{3}a(r) \left[\cos(M\theta)f_6 - \sin(M\theta)f_7\right] \nonumber\\
    &= \frac{\sqrt{5}}{3}b(r)f_1 + \frac{2}{3}a(r) \exp\left(\frac{M\theta}{3}X_\Delta\right)f_6.
    \label{eq:order-three-Higgs-ansatz}
\end{align}
Here \((r,\theta)\) are polar coordinates in the plane transverse to
the vortex.  The profile functions satisfy
\begin{align}
    K(0)=0, \qquad
    a(0)=0, \qquad
    b'(0)=0,
    \label{eq:order-three-origin-boundary-conditions}
\end{align}
and
\begin{align}
    K(r)\longrightarrow1, \qquad
    a(r)\longrightarrow1, \qquad
    b(r)\longrightarrow1 \qquad
    (r\longrightarrow\infty).
    \label{eq:order-three-infinity-boundary-conditions}
\end{align}
Consequently, as \(r\to\infty\),
\begin{align}
    \phi(r,\theta)
    \longrightarrow
    \exp\left(\frac{M\theta}{3}X_\Delta\right)\phi_{\mathrm{vac}},
    \qquad
    \phi_{\mathrm{vac}}
    :=
    \frac{\sqrt{5}}{3}f_1+\frac{2}{3}f_6
    =\sqrt{105} {\mathcal X}{\mathcal Y}{\mathcal Z},
    \label{eq:higgs-vac}
\end{align}
where \(\phi_{\mathrm{vac}}\) is a representative minimum of the Higgs potential that is invariant under \(A_4\)~\cite{Abe:2026zrk}, and the gauge transformation accumulated after one circuit around the vortex is
\begin{align}
    \exp\left(\frac{2\pi\ii M}{3}T_\Delta\right) = r_3^M.
\end{align}

Substituting the ansatz
\eqref{eq:order-three-gauge-ansatz} and
\eqref{eq:order-three-Higgs-ansatz}
into the field equations
\eqref{eq:dimensionless-gauge-equation} and
\eqref{eq:dimensionless-Higgs-equation},
the nontrivial components reduce to
\begin{align}
    0 &= K'' - \frac{1}{r}K' - 4a^2(K-1), \label{eq:order-three-profile-equation-K}\\
    0 &= a'' + \frac{1}{r}a' - \frac{M^2(1-K)^2}{r^2}a - \frac{2\kappa^2}{9} \bigl(-9 + 4a^2 + 5b^2\bigr)a - \frac{10\kappa^2}{9\eta^2}(a^2 - b^2)a, \label{eq:order-three-profile-equation-a}\\
    0 &= b'' + \frac{1}{r}b' - \frac{2\kappa^2}{9}\bigl(-9 + 4a^2 + 5b^2\bigr)b + \frac{8\kappa^2}{9\eta^2}(a^2 - b^2)b. \label{eq:order-three-profile-equation-b}
\end{align}
The tension restricted to this ansatz is
\begin{align}
    E_M[K,a,b] = \frac{4\pi}{9} \int_0^\infty r\,dr\,
    \Biggl[
        &\frac{M^2(K')^2}{4r^2} + (a')^2 + \frac{5}{4}(b')^2 + \frac{M^2(1-K)^2a^2}{r^2} \notag\\
        &+ \frac{\kappa^2}{36}\bigl(-9 + 4a^2 + 5b^2\bigr)^2 + \frac{5\kappa^2}{9\eta^2}(a^2 - b^2)^2
    \Biggr].
\end{align}

At \(\eta=1\), the equation for \(b\) decouples from those for \(K\) and \(a\), and the unique regular finite-tension solution satisfying the boundary conditions is \(b\equiv1\), as shown in Ref.~\cite{Abe:2026zrk}.

At the further critical value \(\kappa=1\), the remaining profile equations \eqref{eq:order-three-profile-equation-K} and \eqref{eq:order-three-profile-equation-a}, together with the restricted tension, reduce to those of the critically coupled Abelian-Higgs\(M\)-vortex in the present normalization \cite{Bogomolny:1975de,Jaffe:1980mj,Abe:2026zrk}.
The restricted tension then admits the Bogomol'nyi completion
\begin{align}
    E_M &= \frac{4\pi}{9} \int_0^\infty r\,dr\,
    \Biggl[\left\{\frac{MK'}{2r} - (1-a^2)\right\}^2 + \left\{a' - \frac{M(1-K)}{r}a\right\}^2\Biggr]
    + \frac{4\pi M}{9} \notag\\
    &\geq \frac{4\pi M}{9}.
    \label{eq:order-three-Bogomolny-completion}
\end{align}
The regular finite-tension solution of Eqs.~\eqref{eq:order-three-profile-equation-K} and \eqref{eq:order-three-profile-equation-a} satisfies the Bogomol'nyi equations
\begin{align}
    a' &= \frac{M(1-K)}{r}a, &
    \frac{MK'}{r} &= 2(1-a^2),
    \label{eq:order-three-Bogomolny-equations}
\end{align}
and therefore saturates the bound,
\begin{align}
    E_M = \frac{4\pi M}{9}.
\end{align}

The inequality above establishes a Bogomol'nyi bound only for field configurations of the form \eqref{eq:order-three-gauge-ansatz} and \eqref{eq:order-three-Higgs-ansatz}.
It does not imply the same lower bound for the full \(SO(3)\) tension, because the full theory also contains gauge and Higgs fluctuations outside this ansatz.
Whether any of these additional fluctuation directions can lower the tension is the stability problem addressed below.

For the stability analysis, we complexify the gauge and Higgs fields.
For the gauge field, we introduce
\begin{align}
    T^0 &:= T_\Delta, &
    T^c &:= \frac{T_1-T_2}{\sqrt{2}}, &
    T^s &:= \frac{T_1+T_2-2T_3}{\sqrt{6}}, \notag\\
    T^+ &:= \frac{T^c+\ii T^s}{\sqrt{2}}, &
    T^- &:= \frac{T^c-\ii T^s}{\sqrt{2}}.
\end{align}
The corresponding generators acting on the Higgs field are
\begin{align}
    X^0 &:= X_\Delta, &
    X^c &:= \frac{X_1-X_2}{\sqrt{2}}, &
    X^s &:= \frac{X_1+X_2-2X_3}{\sqrt{6}}, \notag\\
    X^+ &:= \frac{X^c+\ii X^s}{\sqrt{2}}, &
    X^- &:= \frac{X^c-\ii X^s}{\sqrt{2}}.
\end{align}
For the Higgs field, we define the complex basis
\begin{align}
    u_0 &:= f_1, \notag\\
    u_q &:= \frac{f_{2q}+\ii f_{2q+1}}{\sqrt{2}}, &
    u_{-q} &:= \frac{f_{2q}-\ii f_{2q+1}}{\sqrt{2}},
    \qquad q=1,2,3.
\end{align}
It satisfies
\begin{align}
    X^0u_q = \ii q\,u_q, \qquad
    q = 0,\pm1,\pm2,\pm3,
\end{align}
and, with the complex-bilinear extension of the inner product,
\begin{align}
    \langle u_q,u_{q'}\rangle_\ip = \delta_{q+q',0}.
\end{align}

In the complex basis, the vortex background \eqref{eq:order-three-gauge-ansatz}, \eqref{eq:order-three-Higgs-ansatz} becomes
\begin{align}
    W &= -\frac{M}{3}K(r)T^0\,d\theta, \\
    \phi &= \frac{\sqrt{5}}{3}b(r)u_0 + \frac{2}{3\sqrt{2}}a(r) \left[e^{\ii M\theta}u_3 + e^{-\ii M\theta}u_{-3}\right].
    \label{eq:higgs-bg}
\end{align}
The complexified fluctuations are expanded as
\begin{align}
    B_i &= B_i^0 T^0 + B_i^+ T^+ + B_i^- T^-, \notag\\
    \xi &= \sum_{q=-3}^{3}\xi_q u_q.
\end{align}
We also complexify the spatial vector components by defining
\begin{align}
    B_+^q &:= \frac{B_1^q - \ii B_2^q}{\sqrt{2}}, &
    B_-^q &:= \frac{B_1^q + \ii B_2^q}{\sqrt{2}},
    \qquad q = 0, \pm 1.
\end{align}

We denote the complex-linear extension of the gauge-fixed Hessian \(\widetilde L\) by \(\widetilde{\widetilde L}\).
The complex bases introduced above will be used to decompose \(\widetilde{\widetilde L}\) into angular Fourier blocks.

\subsection{Fourier blocks and symmetry modes}
\label{subsec:order-three-fourier-blocks}
After the complexification introduced in the previous subsection, we write a fluctuation as
\begin{align}
    \Psi=(B_+,B_-,\xi),
\end{align}
where
\begin{align}
    B_\sigma(x) &\in \mathfrak{so}(3)^{\mathbb C}, \qquad \sigma=\pm, \\
    \xi(x) &\in \mathcal H_3^{\mathbb C},
\end{align}
with
\begin{align}
    \mathfrak{so}(3)^{\mathbb C} &:= \operatorname{span}_{\mathbb C}\{T^0,T^+,T^-\}, \notag\\
    \mathcal H_3^{\mathbb C} &:= \operatorname{span}_{\mathbb C}\{u_0,u_{\pm1},u_{\pm2},u_{\pm3}\}.
\end{align}
The complexified fluctuation space is
given by square-integrable fields whose norms converge,
\begin{align}
    \widetilde{\mathcal X} &:= L^2\left(\mathbb R^2; \mathfrak{so}(3)^{\mathbb C} \oplus \mathfrak{so}(3)^{\mathbb C} \oplus \mathcal H_3^{\mathbb C}\right) \notag\\
    &:= \left\{(B_+, B_-, \xi) \, \middle|\, \int_{\mathbb R^2}\left(\|B_+(x)\|^2 + \|B_-(x)\|^2 + \|\xi(x)\|^2\right)d^2x < \infty \right\}.
\end{align}
Here the norms are the Hermitian norms for which the complex bases introduced above are orthonormal.

We now work in polar coordinates,
\begin{align}
    x = r\cos\theta, \qquad
    y = r\sin\theta, \qquad
    d^2x = r\,dr\,d\theta.
\end{align}
Each component of the fluctuation can be expanded in angular Fourier modes,
\begin{align}
    F(r,\theta) = \sum_{m\in\mathbb Z}e^{\ii m\theta}f_m(r),
    \label{eq:order-three-ordinary-Fourier-expansion}
\end{align}
with
\begin{align}
    f_m \in\mathscr H_r, \qquad
    \mathscr H_r:=L^2\bigl((0,\infty),r\,dr\bigr) = \left\{f : (0,\infty) \longrightarrow \mathbb C \, \middle|\, \int_0^\infty \|f(r)\|^2 r\,dr < \infty \right\}.
\end{align}
For later use, we denote the subspace with fixed angular dependence by
\begin{align}
    e^{\ii m\theta}\mathscr H_r :=\left\{e^{\ii m\theta}f(r)\,\middle|\,f\in\mathscr H_r\right\}.
\end{align}

In the present problem, the fluctuation has thirteen complex scalar components:
six gauge-field components and seven Higgs-field components,
\begin{align}
    \bigl(
        B_+^0,\,
        B_-^0,\,
        B_+^+,\,
        B_-^+,\,
        B_+^-,\,
        B_-^-,\,
        \xi_0,\,
        \xi_{+1},\,
        \xi_{-1},\,
        \xi_{+2},\,
        \xi_{-2},\,
        \xi_{+3},\,
        \xi_{-3}
    \bigr).
    \label{eq:order-three-thirteen-components}
\end{align}
Each of these components separately admits an expansion of the form
\eqref{eq:order-three-ordinary-Fourier-expansion}.
For the stability analysis, we organize the Fourier modes into three
classes, labeled by \(\rho=0,\pm1\). Since the Higgs background (\ref{eq:higgs-bg}) contains
the components \(e^{\pm iM\theta}u_{\pm3}\), fluctuations whose
\(X_0\) charges differ by three can mix, so that \(\rho\) labels the
\(X_0\) charge modulo three:
\begin{equation}
    \widetilde{\mathcal X}
    =
    \bigoplus_{\ell\in\mathbb Z}
    \bigoplus_{\rho\in\{0,\pm1\}}
    \mathcal F^\rho_\ell .
\end{equation}
The corresponding angular Fourier-index shifts associated with these modulo-three charge classes are fixed by the background dependence \(e^{\pm iM\theta}\), giving shifts by \(\pm M\) for the Higgs components. 
In addition, the spatially complexified gauge
components \(B_\pm\) carry angular helicity \(\pm1\), leading to the
shifts \(\ell\mp1\). Using the radial space \(\mathscr H_r\) defined
above, the three Fourier blocks are therefore defined by
\begin{align}
    \mathcal F_\ell^0 &:=
    \underbrace{e^{\ii(\ell-1)\theta}\mathscr H_r}_{B_+^0} \oplus
    \underbrace{e^{\ii(\ell+1)\theta}\mathscr H_r}_{B_-^0}\oplus
    \underbrace{\ii e^{\ii\ell\theta}\mathscr H_r}_{\xi_0} \oplus
    \underbrace{\ii e^{\ii(\ell+M)\theta}\mathscr H_r}_{\xi_{+3}} \oplus
    \underbrace{-\ii e^{\ii(\ell-M)\theta}\mathscr H_r}_{\xi_{-3}},
    \\[1ex]
    \mathcal F_\ell^1 &:=
    \underbrace{e^{\ii(\ell-1)\theta}\mathscr H_r}_{B_+^+} \oplus
    \underbrace{e^{\ii(\ell+1)\theta}\mathscr H_r}_{B_-^+} \oplus
    \underbrace{e^{\ii\ell\theta}\mathscr H_r}_{\xi_{+1}} \oplus
    \underbrace{e^{\ii(\ell-M)\theta}\mathscr H_r}_{\xi_{-2}},
    \\[1ex]
    \mathcal F_\ell^{-1} &:=
    \underbrace{e^{\ii(\ell-1)\theta}\mathscr H_r}_{B_+^-} \oplus
    \underbrace{e^{\ii(\ell+1)\theta}\mathscr H_r}_{B_-^-} \oplus
    \underbrace{e^{\ii(\ell+M)\theta}\mathscr H_r}_{\xi_{+2}} \oplus
    \underbrace{e^{\ii\ell\theta}\mathscr H_r}_{\xi_{-1}}.
\end{align}
Here, omitted components are understood to vanish.  
The constant phases \(\pm \ii\) in the above equations are chosen so that the corresponding radial operators have real coefficients.
For example, a general element of \(\mathcal F_\ell^0\) is written as
\begin{align}
    \begin{pmatrix}
        B_+^0\\
        B_-^0\\
        B_+^+\\
        B_-^+\\
        B_+^-\\
        B_-^-\\
        \xi_0\\
        \xi_{+1}\\
        \xi_{-1}\\
        \xi_{+2}\\
        \xi_{-2}\\
        \xi_{+3}\\
        \xi_{-3}
    \end{pmatrix}
    =
    \begin{pmatrix}
        e^{\ii(\ell-1)\theta}S_+^0(r)\\
        e^{\ii(\ell+1)\theta}S_-^0(r)\\
        0\\
        0\\
        0\\
        0\\
        \ii e^{\ii\ell\theta}\chi_0(r)\\
        0\\
        0\\
        0\\
        0\\
        \ii e^{\ii(\ell+M)\theta}\chi_{+3}(r)\\
        -\ii e^{\ii(\ell-M)\theta}\chi_{-3}(r)
    \end{pmatrix}
    \in \mathcal F_\ell^0 ,
    \label{eq:order-three-Fourier-block-example}
\end{align}
for some \(S_+^0,\, S_-^0,\, \chi_0,\, \chi_{+3},\, \chi_{-3} \in\mathscr H_r\).

By explicitly substituting the Fourier expansions above into the complexified gauge-fixed Hessian and evaluating its components, we find that subspaces with different labels \((\rho,\ell)\) do not mix:
\begin{align}
    \widetilde{\widetilde L}\,\mathcal F_\ell^\rho
    \subset \mathcal F_\ell^\rho, \qquad
    \ell\in\mathbb Z, \qquad
    \rho=0, \pm1.
\end{align}

To extract the corresponding radial operators, we collect the nonvanishing components of a mode in \(\mathcal F_\ell^\rho\) and factor out their fixed angular dependencies:
\begin{align}
    \Psi_\ell^\rho(r,\theta)
    = U_\ell^\rho(\theta)\,u_\ell^\rho(r).
\end{align}
Collecting the nonvanishing components of each Fourier block, we define the radial coefficient vectors
\begin{align}
    u_\ell^0(r) &:=
    \begin{pmatrix}
        S_+^0(r)\\
        S_-^0(r)\\
        \chi_0(r)\\
        \chi_{+3}(r)\\
        \chi_{-3}(r)
    \end{pmatrix}, &
    u_\ell^1(r) &:=
    \begin{pmatrix}
        S_+^+(r)\\
        S_-^+(r)\\
        \chi_{+1}(r)\\
        \chi_{-2}(r)
    \end{pmatrix}, &
    u_\ell^{-1}(r) &:=
    \begin{pmatrix}
        S_+^-(r)\\
        S_-^-(r)\\
        \chi_{+2}(r)\\
        \chi_{-1}(r)
    \end{pmatrix},
\end{align}
and the diagonal matrices containing the prescribed angular factors and phase conventions are
\begin{align}
    U_\ell^0 &:= \operatorname{diag}\left(
        e^{\ii(\ell-1)\theta},
        e^{\ii(\ell+1)\theta},
        \ii e^{\ii\ell\theta},
        \ii e^{\ii(\ell+M)\theta},
        -\ii e^{\ii(\ell-M)\theta}
    \right), \notag\\
    U_\ell^1 &:= \operatorname{diag}\left(
        e^{\ii(\ell-1)\theta},
        e^{\ii(\ell+1)\theta},
        e^{\ii\ell\theta},
        e^{\ii(\ell-M)\theta}
    \right), \notag\\
    U_\ell^{-1} &:= \operatorname{diag}\left(
        e^{\ii(\ell-1)\theta},
        e^{\ii(\ell+1)\theta},
        e^{\ii(\ell+M)\theta},
        e^{\ii\ell\theta}
    \right).
\end{align}
Because \(\widetilde{\widetilde L}\) preserves each Fourier subspace, its action retains the same angular factors and can be written as
\begin{align}
    \widetilde{\widetilde L}\left(U_\ell^\rho u_\ell^\rho\right)
    = U_\ell^\rho \left(L_\ell^\rho u_\ell^\rho\right).
    \label{eq:order-three-radial-block-definition}
\end{align}
This relation defines \(L_\ell^\rho\) as a differential operator acting only on the radial coefficient vector \(u_\ell^\rho(r)\).
Since \(u_\ell^0\) has five components, whereas \(u_\ell^1\) and \(u_\ell^{-1}\) each have four, \(L_\ell^0\) is a \(5\times5\) matrix-valued differential operator, while \(L_\ell^1\) and \(L_\ell^{-1}\) are \(4\times4\) ones.
Each radial block takes the Schrödinger-type form
\begin{align}
    L_\ell^\rho = -\Delta_r\,\mathbf 1 + V_\ell^\rho(r), \qquad
    \Delta_r := \frac{d^2}{dr^2} + \frac{1}{r}\frac{d}{dr},
\end{align}
where \(V_\ell^\rho(r)\) is a real matrix-valued function of size \(5\times5\) for \(\rho=0\) and \(4\times4\) for \(\rho=\pm1\).
The explicit matrices \(V_\ell^\rho(r)\) are given in Appendix~\ref{app:order-three-radial-blocks}.
The reality of these radial operators reflects the fact that the gauge-fixed Hessian before complexification is real.
Consequently, its complexification commutes with complex conjugation:
\begin{align}
    \widetilde{\widetilde L}\,\overline{\Psi}
    = \overline{\widetilde{\widetilde L}\Psi}.
    \label{eq:order-three-Hessian-reality}
\end{align}
To see how this property relates blocks with different Fourier labels, consider the general element of \(\mathcal F_\ell^0\) displayed in Eq.~\eqref{eq:order-three-Fourier-block-example}.
After taking its complex conjugate and re-expanding it in the same complex bases, the five nonvanishing components \((B_+^0,B_-^0,\xi_0,\xi_{+3},\xi_{-3})\) become
\begin{align}
    \begin{pmatrix}
        e^{\ii(-\ell-1)\theta}\overline{S_-^0(r)}\\
        e^{\ii(-\ell+1)\theta}\overline{S_+^0(r)}\\
        \ii e^{-\ii\ell\theta}
        \bigl[-\overline{\chi_0(r)}\bigr]\\
        \ii e^{\ii(-\ell+M)\theta}\overline{\chi_{-3}(r)}\\
        -\ii e^{\ii(-\ell-M)\theta}\overline{\chi_{+3}(r)}
    \end{pmatrix}.
\end{align}
This has precisely the angular dependence of a mode in \(\mathcal F_{-\ell}^0\).
Thus, complex conjugation maps its radial coefficient vector according to
\begin{align}
    \begin{pmatrix}
        S_+^0\\
        S_-^0\\
        \chi_0\\
        \chi_{+3}\\
        \chi_{-3}
    \end{pmatrix}
    \longmapsto
    P_0
    \begin{pmatrix}
        \overline{S_+^0}\\
        \overline{S_-^0}\\
        \overline{\chi_0}\\
        \overline{\chi_{+3}}\\
        \overline{\chi_{-3}}
    \end{pmatrix},
    \qquad
    P_0 :=
    \begin{pmatrix}
        0&1&0&0&0\\
        1&0&0&0&0\\
        0&0&-1&0&0\\
        0&0&0&0&1\\
        0&0&0&1&0
    \end{pmatrix}.
\end{align}
Note that \(P_0\) converts the complex-conjugated radial vector from
\(\mathcal F_\ell^0\) to the standard basis of \(\mathcal F_{-\ell}^0\).
Let \(v\) denote the radial coefficient vector before complex conjugation.
Applying the Hessian before or after complex conjugation must give the same result by Eq.~\eqref{eq:order-three-Hessian-reality}.
Therefore,
\begin{align}
    L_{-\ell}^0P_0\overline{v}
    = P_0\overline{L_\ell^0v}
    = P_0 L_\ell^0\overline{v},
\end{align}
where the second equality follows from the reality of \(L_\ell^0\).
Since \(v\) is arbitrary, this gives
\begin{align}
    P_0 L_\ell^0 P_0^{-1} = L_{-\ell}^0.
\end{align}
The same argument maps \(\mathcal F_\ell^1\) to \(\mathcal F_{-\ell}^{-1}\) and gives
\begin{align}
    PL_\ell^1P^{-1} &= L_{-\ell}^{-1}, &
    P &:=
    \begin{pmatrix}
        0&1&0&0\\
        1&0&0&0\\
        0&0&0&1\\
        0&0&1&0
    \end{pmatrix}.
    \label{eq:order-three-charged-block-similarity}
\end{align}
These relations can also be verified directly from the explicit matrix elements of \(V_\ell^0\), \(V_\ell^1\) and \(V_{-\ell}^{-1}\) listed in Appendix~\ref{app:order-three-radial-blocks}.
Hence \(L_\ell^0\) and \(L_{-\ell}^0\) have the same eigenvalues, and so do \(L_\ell^1\) and \(L_{-\ell}^{-1}\).
It is therefore sufficient to analyze
\begin{align}
    \rho=0,\quad \ell\geq0,
    \qquad\text{and}\qquad
    \rho=1,\quad \ell\in\mathbb Z.
\end{align}

We next identify the zero modes corresponding to infinitesimal translations of the vortex center.
An infinitesimal translation in the \(x^j\) direction is described by
\begin{align}
    S_j := \begin{pmatrix}
        \partial_j W_1\\
        \partial_j W_2\\
        \partial_j\phi
    \end{pmatrix}
    = 
    \begin{pmatrix}
        \partial_j W_i\\
        \partial_j\phi
    \end{pmatrix},
    \qquad j=1,2.
\end{align}
Although \(S_j\) is a zero mode of the ungauge-fixed Hessian by translational invariance, it does not in general satisfy the gauge condition \eqref{eq:background-gauge-condition}, and hence need not be a zero mode of the gauge-fixed Hessian \(\widetilde L\).
For a fixed translation direction \(j\), we choose the gauge parameter \(\alpha=W_j\).
The corresponding gauge zero mode is
\begin{align}
    R W_j = \begin{pmatrix}
        -\mathcal D_i W_j\\
        W_j^a X_a\phi
    \end{pmatrix}.
\end{align}
The resulting fluctuation is
\begin{align}
    Z_j &:= S_j + R W_j \notag\\
    &= \begin{pmatrix}
        \partial_j W_i - \mathcal D_i W_j\\
        \partial_j \phi + W_j^a X_a\phi
    \end{pmatrix}
    = \begin{pmatrix}
        G_{ji}\\
        D_j \phi
    \end{pmatrix}.
    \label{eq:order-three-covariant-translation-mode}
\end{align}
Since the added term \(RW_j\) is a gauge zero mode, \(Z_j\) represents the same infinitesimal translation of the vortex as \(S_j\).
Moreover, \(Z_j\) satisfies the gauge condition \eqref{eq:background-gauge-condition}.
Indeed, the gauge-field equation \eqref{eq:dimensionless-gauge-equation} gives
\begin{align}
    \Gamma^a(Z_j) = -(\mathcal D_i^\dagger G_{ij})^a - \left\langle X_a\phi, D_j\phi\right\rangle_\ip = 0.
\end{align}
Since \(Z_j\) is also a zero mode of the ungauge-fixed Hessian, these relations imply
\begin{align}
    \widetilde L Z_j = 0, \qquad j = 1,2.
\end{align}
Thus, \(Z_j\) is a zero mode of the gauge-fixed Hessian corresponding to an infinitesimal translation of the vortex center in the \(x^j\) direction.

To determine which Fourier blocks contain these modes, we combine the two real translational zero modes as
\begin{align}
    Z^{(+)} &:= Z_1 + \ii Z_2, &
    Z^{(-)} &:= Z_1 - \ii Z_2.
\end{align}
For convenience, we introduce
\begin{align}
    \mathcal B_\Delta &:= -\frac{M}{3r}K', &
    \mathcal A_\pm &:= a' \pm \frac{M(1-K)}{r}a.
\end{align}
Substituting the order-three background fields \eqref{eq:order-three-gauge-ansatz} and \eqref{eq:order-three-Higgs-ansatz} into \eqref{eq:order-three-covariant-translation-mode}, and expressing the result in the complex bases introduced above, gives
\begin{align}
    Z^{(+)} &=
    \begin{pmatrix}
        -\sqrt{2}\ii\mathcal B_\Delta\\
        0\\
        0\\
        0\\
        0\\
        0\\
        \dfrac{\sqrt{5}}{3}b'e^{\ii\theta}\\
        0\\
        0\\
        0\\
        0\\
        \dfrac{\sqrt{2}}{3}\mathcal A_-e^{\ii(M+1)\theta}\\
        \dfrac{\sqrt{2}}{3}\mathcal A_+e^{\ii(1-M)\theta}
    \end{pmatrix}
    \in \mathcal F_1^0, &
    Z^{(-)} &=
    \begin{pmatrix}
        0\\
        \sqrt{2}\ii\mathcal B_\Delta\\
        0\\
        0\\
        0\\
        0\\
        \dfrac{\sqrt{5}}{3}b'e^{-\ii\theta}\\
        0\\
        0\\
        0\\
        0\\
        \dfrac{\sqrt{2}}{3}\mathcal A_+e^{\ii(M-1)\theta}\\
        \dfrac{\sqrt{2}}{3}\mathcal A_-e^{-\ii(M+1)\theta}
    \end{pmatrix}
    \in \mathcal F_{-1}^0,
\end{align}
where the component order is that of \eqref{eq:order-three-thirteen-components}.
Since \(Z_1\) and \(Z_2\) are real fluctuations, \(Z^{(-)}=\overline{Z^{(+)}}\).
Thus, the modes in \(\mathcal F_1^0\) and \(\mathcal F_{-1}^0\) are exchanged by the complex-conjugation map represented by \(P_0\).

We finally consider the infinitesimal deformation generated by a spatial rotation of the vortex.
Since the background gauge field \eqref{eq:order-three-gauge-ansatz} is rotationally symmetric, its variation vanishes.
The raw rotational mode is therefore
\begin{align}
    Z_{\mathrm{rot}}
    &:= - \begin{pmatrix}
        0\\
        \partial_\theta\phi
    \end{pmatrix}.
\end{align}
Substituting the order-three Higgs field ansatz \eqref{eq:order-three-Higgs-ansatz}, we find
\begin{align}
    \partial_\theta\phi
    &= -\frac{2M}{3}a(r) \left[\sin(M\theta)f_6 + \cos(M\theta)f_7\right]
    = \frac{M}{3}X^0\phi.
\end{align}
Hence
\begin{align}
    Z_{\mathrm{rot}}
    = \begin{pmatrix}
        0\\[1mm]
        -\dfrac{M}{3}X^0\phi
    \end{pmatrix}.
\end{align}

Now consider the constant gauge parameter
\begin{align}
    \alpha_{\mathrm{rot}} := -\frac{M}{3}T^0.
\end{align}
The corresponding infinitesimal gauge transformation is
\begin{align}
    R\alpha_{\mathrm{rot}} &= \begin{pmatrix}
        -\mathcal D_i\alpha_{\mathrm{rot}}\\
        \alpha_{\mathrm{rot}}^aX_a\phi
    \end{pmatrix}
    = \begin{pmatrix}
        0\\[1mm]
        -\dfrac{M}{3}X^0\phi
    \end{pmatrix}
    = Z_{\mathrm{rot}}.
\end{align}
Thus, the infinitesimal rotational deformation is a gauge zero mode rather than an additional physical zero mode, and is lifted by the background gauge fixing.

\subsection{Numerical results}
\label{subsec:order-three-numerical-result}
We numerically solved the radial equations \eqref{eq:order-three-profile-equation-K}--\eqref{eq:order-three-profile-equation-b} and substituted the resulting profiles into the radial operators \(L_\ell^\rho\) in  Appendix~\ref{app:order-three-radial-blocks}.
For each Fourier block, we then numerically solved the radial eigenvalue problem
\begin{align}
    L_\ell^\rho u_\ell^\rho = \lambda\,u_\ell^\rho.
    \label{eq:order-three-radial-eigenvalue-problem}
\end{align}
As explained in Sec.~\ref{subsec:order-three-fourier-blocks}, the blocks \((0,\ell)\) and \((0,-\ell)\) have the same spectrum, and likewise \((1,\ell)\) and \((-1,-\ell)\) have the same spectrum.
Accordingly, in the numerical scan we use \(\rho=0\) with \(\ell\geq0\) and \(\rho=1\) with \(\ell\in\mathbb Z\) as independent representatives.

The translational zero modes belong to the \((\rho,\ell)=(0,\pm1)\) blocks, as shown in Sec.~\ref{subsec:order-three-fourier-blocks}.
In these blocks, the numerical eigenvalue closest to zero is identified with the translational zero mode.
Let \(\lambda_{\rho,\ell}^{(M)}(\kappa,\eta)\) denote the lowest eigenvalue of \(L_\ell^\rho\), excluding the translational zero mode, and define
\begin{align}
    \lambda_M(\kappa,\eta)
    := \min_{\substack{\rho\in\{0,\pm1\}\\ \ell\in\mathbb Z}}
    \lambda_{\rho,\ell}^{(M)}(\kappa,\eta).
\end{align}

\begin{figure}[t]
    \centering
    \includegraphics[width=1\textwidth]{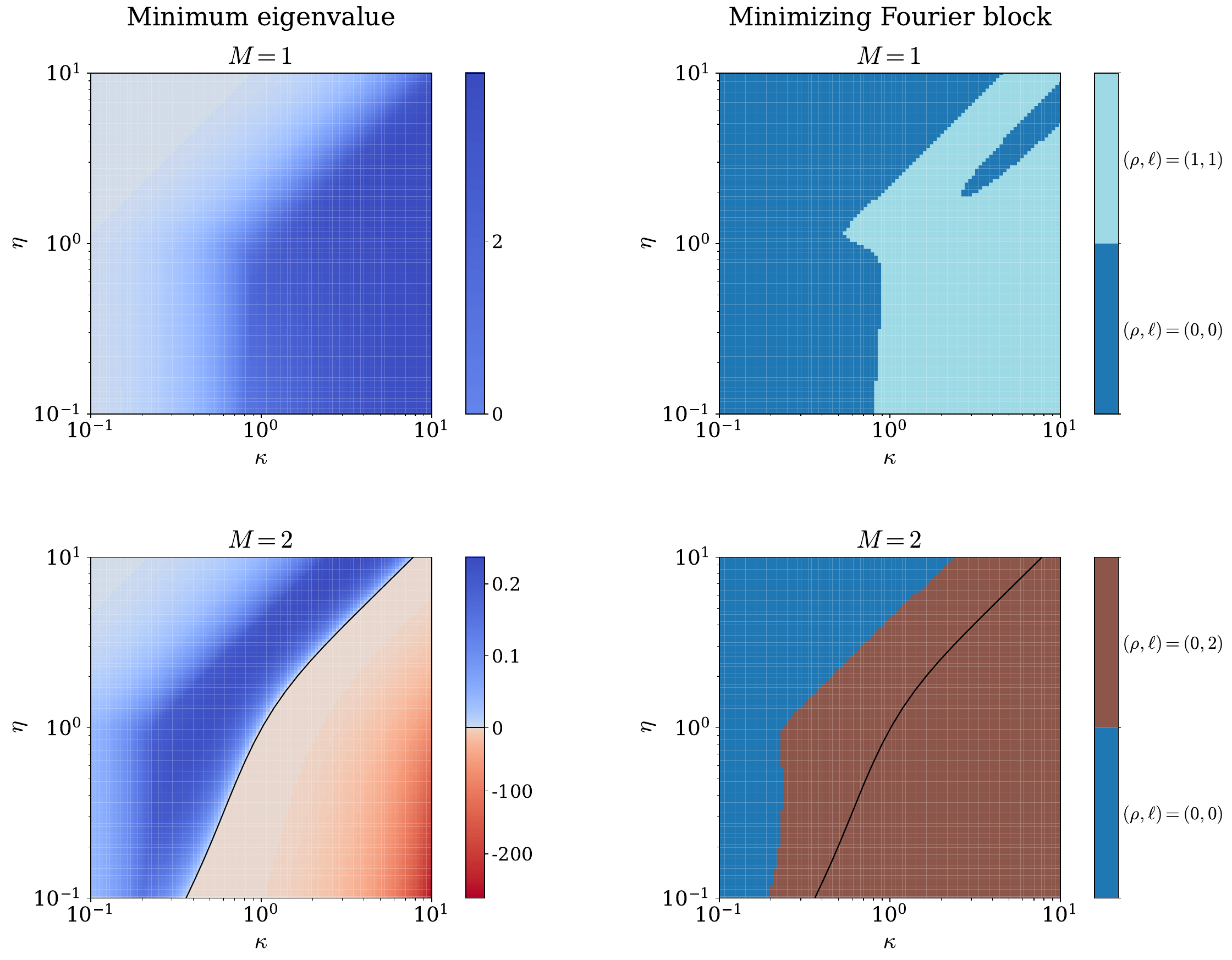}
        
    \caption{
        Numerical stability analysis of the order-three vortex family for \(M=1\) and \(2\).
        In each row, the left panel shows the minimum eigenvalue \(\lambda_M(\kappa,\eta)\) of the gauge-fixed Hessian, with the translational zero mode excluded, and the right panel shows the representative Fourier block \((\rho,\ell)\) in which this minimum is attained.
        The sector labels use \(\rho=0\) with \(\ell\geq0\) and \(\rho=1\) with \(\ell\in\mathbb Z\).
        The black curves indicate \(\lambda_M(\kappa,\eta)=0\).
    }
    \label{fig:order-three-numerics}
\end{figure}

Figure~\ref{fig:order-three-numerics} shows the numerical values of \(\lambda_M(\kappa,\eta)\) for \(M=1\) and \(2\), obtained by solving Eq.~\eqref{eq:order-three-radial-eigenvalue-problem}.
The left column shows the minimum eigenvalue \(\lambda_M(\kappa,\eta)\), while the right column identifies the representative Fourier block in which this minimum is attained.
The sector labels in the right column use \((0,\ell)\) with \(\ell\geq0\) and \((1,\ell)\) with \(\ell\in\mathbb Z\);
the symmetry-related blocks \((0,-\ell)\) and \((-1,-\ell)\) have the same spectra as \((0,\ell)\) and \((1,\ell)\), respectively.

For \(M=1\), no negative eigenvalue is found in the parameter region shown.
For \(M=2\), by contrast, a region with \(\lambda_2(\kappa,\eta)<0\) appears.
The negative mode in this region lies in the \((\rho,\ell)=(0,2)\) block.
At the linearized level, this mode is naturally interpreted as an infinitesimal deformation associated with splitting the \(M=2\) vortex into two \(M=1\) order-three vortices.
This interpretation is also suggested by the Fourier-block structure.
The translational modes belong to the \(\rho=0\) sector, and the
splitting deformation corresponds to a relative displacement of the
two vortex cores. Its quadrupolar angular dependence then places it in
the \(\ell=2\) mode.

The special point \((\kappa,\eta)=(1,1)\) is treated analytically in Sec.~\ref{subsec:order-three-analytical-stability}, where we prove for arbitrary \(M\) that the gauge-fixed Hessian is non-negative and determine its zero-mode structure.

\begin{figure}[t]
    \centering
    \includegraphics[width=0.72\textwidth]{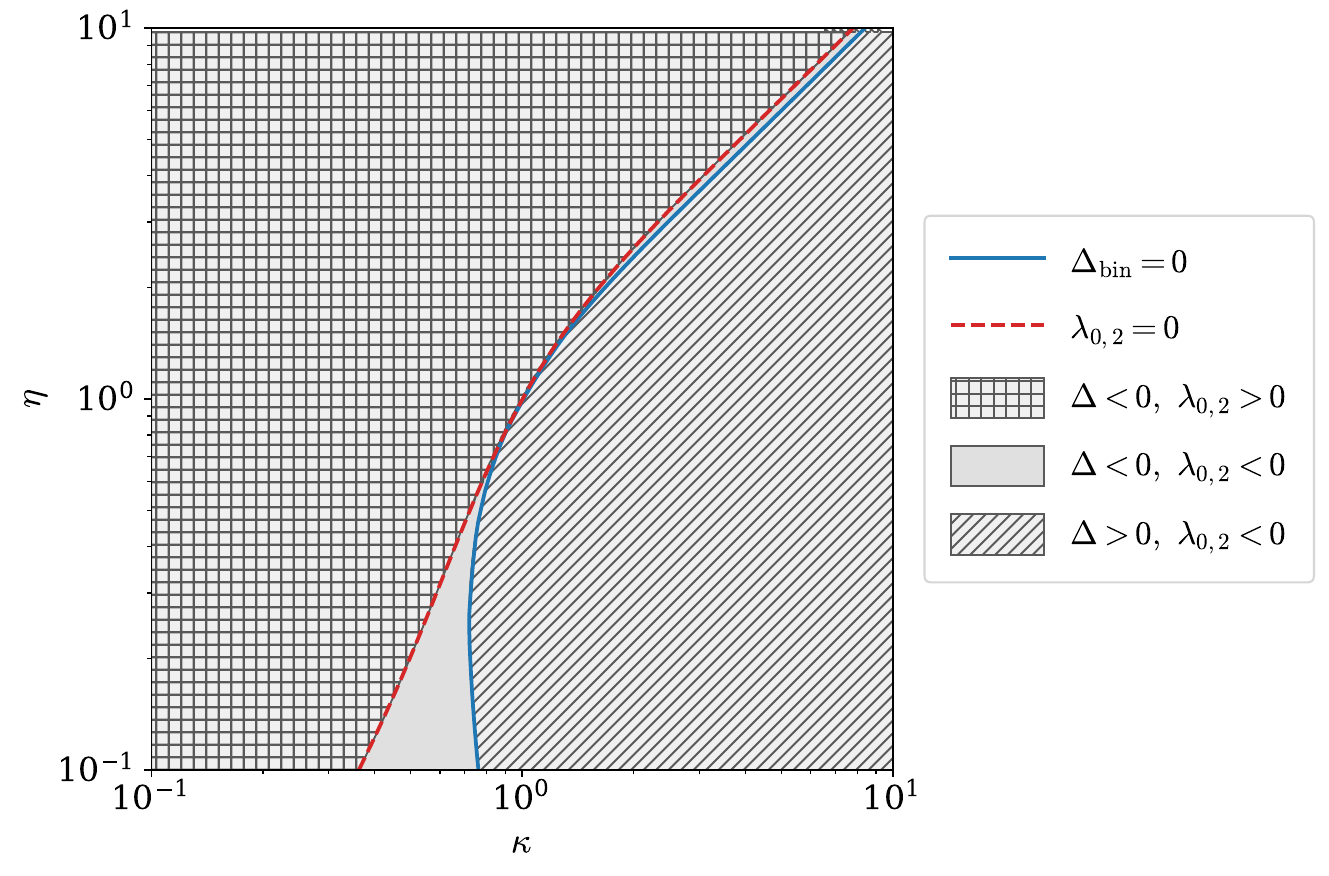}
    \caption{
        Comparison of the energy difference \(\Delta_{\mathrm{bin}} = E_2 - 2E_1\) and the lowest eigenvalue \(\lambda_{0,2}^{(2)}\) in the \((\rho,\ell)=(0,2)\) Fourier block for the \(M=2\) order-three vortex.
        The solid blue curve indicates \(\Delta_{\mathrm{bin}}=0\), and the dashed red curve indicates \(\lambda_{0,2}^{(2)}=0\).
        The differently shaded regions indicate the three sign combinations of \(\Delta_{\mathrm{bin}}\) and \(\lambda_{0,2}^{(2)}\) found in the displayed parameter range.
    }
    \label{fig:order-three-binding-vs-stability-M2}
\end{figure}

For \(M=2\), we further compare the lowest eigenvalue in the \((\rho,\ell)=(0,2)\) Fourier block with the energy difference between the axisymmetric \(M=2\) vortex and two infinitely separated \(M=1\) vortices.
Let \(E_M(\kappa,\eta)\) denote the tension of the axisymmetric \(M\)-vortex solution, and define
\begin{align}
    \Delta_{\mathrm{bin}}(\kappa,\eta) := E_2(\kappa,\eta) - 2E_1(\kappa,\eta).
\end{align}
Thus, \(\Delta_{\mathrm{bin}}<0\) means that the axisymmetric \(M=2\) vortex has lower tension than two infinitely separated \(M=1\) vortices.
Figure~\ref{fig:order-three-binding-vs-stability-M2} compares \(\Delta_{\mathrm{bin}}\) with \(\lambda_{0,2}^{(2)}\).
We compare these quantities because the \((\rho,\ell)=(0,2)\) block contains the infinitesimal splitting deformation of the axisymmetric \(M=2\) vortex into two \(M=1\) vortices.
The two curves do not coincide, and there is a parameter region in which
\begin{align}
    \Delta_{\mathrm{bin}} < 0, \qquad
    \lambda_{0,2}^{(2)} < 0.
\end{align}
In this region, the axisymmetric \(M=2\) vortex solution has lower tension than two infinitely separated \(M=1\) vortices, but is unstable against the splitting deformation represented by the \((\rho,\ell)=(0,2)\) mode.
If this negative mode develops into a separation of two \(M=1\) vortices, the energy decreases as the vortices begin to separate from the axisymmetric configuration.
However, their energy at infinite separation is higher because \(\Delta_{\mathrm{bin}}<0\).
Together, these observations suggest that the energy as a function of the vortex separation may have a minimum at a finite separation, corresponding to a molecular bound state of two parallel \(M=1\) vortices.
For completeness, the remaining sign combinations admit a simpler
interpretation. If \(\Delta_{\rm bin}<0\) while
\(\lambda^{(2)}_{0,2}>0\), the axisymmetric \(M=2\) vortex lies below
the energy of two infinitely separated \(M=1\) vortices and is locally
stable against splitting. By contrast, if \(\Delta_{\rm bin}>0\) and
\(\lambda^{(2)}_{0,2}<0\), the axisymmetric configuration is unstable
to splitting and lies above the energy of two infinitely separated \(M=1\) vortices.

\subsection{Stability at \texorpdfstring{\(\kappa=\eta=1\)}{kappa=eta=1}}
\label{subsec:order-three-analytical-stability}
We now prove the linear stability of the order-three vortex at \(\kappa=\eta=1\) for arbitrary \(M\geq1\).
As discussed in Sec.~\ref{subsec:order-three-ansatz-complex-basis},
the regular finite-tension solution has \(b=1\) \cite{Abe:2026zrk}, and the remaining profiles satisfy the Bogomol'nyi equations \eqref{eq:order-three-Bogomolny-equations} \cite{Bogomolny:1975de,Jaffe:1980mj,Abe:2026zrk}.
The profile \(a\) further satisfies
\begin{align}
    0 \leq a(r) \leq 1, \qquad r > 0,
    \label{eq:order-three-BPS-profile-bound}
\end{align}
as shown in Appendix~\ref{app:order-three-BPS-profile-bounds}.

We show that the gauge-fixed Hessian is non-negative and has precisely \(2M\) zero modes:
\begin{align}
    \widetilde L &\geq 0, &
    \dim_{\mathbb R}\ker\widetilde L &= 2M.
    \label{eq:order-three-BPS-stability-result}
\end{align}
The \(2M\) zero modes correspond to variations of the positions of the \(M\) vortex cores in the transverse plane.
In particular, for \(M=1\), the only zero modes are the two translational modes. Together with the absence of negative modes, this shows that the vortex is linearly stable.

We first consider the \(\rho=0\) sector.
At \(\kappa=\eta=1\), using \(b=1\) and the Bogomol'nyi equations \eqref{eq:order-three-Bogomolny-equations}, the radial operator \(L_\ell^0\), whose explicit matrix form is given in Appendix~\ref{app:order-three-radial-blocks}, decomposes into a one-component operator and two \(2\times2\) operators related by \(\ell\leftrightarrow-\ell\):
\begin{align}
    L_\ell^0 \simeq H_{\ell,M}^{\mathrm{AH}} \oplus H_{-\ell,M}^{\mathrm{AH}} \oplus H_\ell^{\mathrm{neu}}.
    \label{eq:order-three-BPS-rho0-decomposition}
\end{align}
Here \(H_{\ell,M}^{\mathrm{AH}}\) and \(H_{-\ell,M}^{\mathrm{AH}}\) act on \((S_+^0,\chi_{-3})\) and \((S_-^0,\chi_{+3})\), respectively, while \(H_\ell^{\mathrm{neu}}\) acts on \(\chi_0\).
Their explicit forms are
\begin{align}
    H_{\ell,M}^{\mathrm{AH}} &=
    \begin{pmatrix}
        -\Delta_r + \dfrac{(\ell-1)^2}{r^2} + 4a^2 & -4a' \\[3mm]
        -4a' & -\Delta_r + \dfrac{\bigl[\ell-M(1-K)\bigr]^2}{r^2} + 6a^2 - 2
    \end{pmatrix}, \\
    H_\ell^{\mathrm{neu}} &=
    -\Delta_r + \frac{\ell^2}{r^2} + 4.
\end{align}

Defining
\begin{align}
    \mathcal A_{\ell,M}^{\mathrm{AH}} &:=
    \begin{pmatrix}
        \dfrac{d}{dr} - \dfrac{\ell-1}{r} & 2a \\[3mm]
        -2a & -\dfrac{d}{dr} - \dfrac{\ell-M(1-K)}{r}
    \end{pmatrix}, \\
    d_\ell &:= \frac{d}{dr} - \frac{\ell}{r},
\end{align}
the operators in Eq.~\eqref{eq:order-three-BPS-rho0-decomposition} factorize as
\begin{align}
    H_{\pm\ell,M}^{\mathrm{AH}} &= \left(\mathcal A_{\pm\ell,M}^{\mathrm{AH}}\right)^\dagger \mathcal A_{\pm\ell,M}^{\mathrm{AH}}, &
    H_\ell^{\mathrm{neu}} &= d_\ell^\dagger d_\ell + 4.
\end{align}
Here \(\dagger\) denotes the formal adjoint with respect to the radial measure \(r\,dr\):
for an operator \(A\) acting on radial functions, \(A^\dagger\) is defined by
\begin{align}
    \int_0^\infty \left(\overline{f} A g\right) r\,dr = \int_0^\infty \left(\overline{(A^\dagger f)} g\right) r\,dr
\end{align}
for \(f\) and \(g\) with vanishing boundary terms.
In particular,
\begin{align}
    \left(\frac{d}{dr}\right)^\dagger &= -\frac{d}{dr} - \frac{1}{r}, &
    d_\ell^\dagger &= -\frac{d}{dr} - \frac{\ell+1}{r}.
\end{align}
With this convention, the factorized forms above are manifestly non-negative.
It follows that the entire \(\rho=0\) sector is non-negative.

After a constant rescaling associated with the normalization and charge conventions of the present model,
\(H_{\ell,M}^{\mathrm{AH}}\) and \(H_{-\ell,M}^{\mathrm{AH}}\) have the same form as the gauge-fixed radial fluctuation operators of a critically coupled Abelian-Higgs \(M\)-vortex \cite{GustafsonSigal2000}.
For this operator, the zero modes are known to be precisely those generated by varying the positions of the \(M\) vortex cores, giving \(2M\) real zero modes \cite{Stuart:1994yc,GustafsonSigal2000}.
The \(\rho=0\) sector therefore contains precisely these \(2M\) vortex-core position zero modes.

We next consider the \(\rho=1\) sector.
At \(\kappa=\eta=1\), using \(b=1\) and the Bogomol'nyi equations \eqref{eq:order-three-Bogomolny-equations}, the radial operator \(L_\ell^1\), whose explicit matrix form is given in Appendix~\ref{app:order-three-radial-blocks}, decomposes as
\begin{align}
    L_\ell^1 \simeq H_{\ell,M}^{(S)} \oplus H_{\ell,M}^{(\chi)} \oplus H_{\ell,M}^{\perp}.
\end{align}
The three operators act on \(S_-^+\), \(\chi_{+1}\), and \((S_+^+,\chi_{-2})\), respectively.
They are
\begin{align}
    H_{\ell,M}^{(S)} &= -\Delta_r + \frac{\bigl[3(\ell+1)-MK\bigr]^2}{9r^2} + \frac{2}{3}(7 - a^2), \\
    H_{\ell,M}^{(\chi)} &= -\Delta_r + \frac{(3\ell-MK)^2}{9r^2} + \frac{2}{3}(a^2+5), \\
    H_{\ell,M}^{\perp} &=
    \begin{pmatrix}
        -\Delta_r+p_{\ell,M}^2+2(a^2+1) & \dfrac{4}{\sqrt{3}}a' \\[2mm]
        \dfrac{4}{\sqrt{3}}a' & -\Delta_r+q_{\ell,M}^2+\dfrac{4}{3}(2a^2+1)
    \end{pmatrix},
\end{align}
where
\begin{align}
    p_{\ell,M} &:= \frac{3(\ell-1)-MK}{3r},\\
    q_{\ell,M} &:= \frac{3(\ell-M)+2MK}{3r}.
\end{align}
Using \eqref{eq:order-three-BPS-profile-bound}, 
together with the non-negativity of the radial kinetic and squared angular terms,
we obtain
\begin{align}
    H_{\ell,M}^{(S)} &\geq 4, &
    H_{\ell,M}^{(\chi)} &\geq \frac{10}{3}.
    \label{eq:order-three-BPS-decoupled-lower-bounds}
\end{align}

For the remaining two-component operator \(H_{\ell,M}^{\perp}\), we introduce
\begin{align}
    \mathcal A_{\ell,M}^{\perp} :=
    \begin{pmatrix}
        \dfrac{d}{dr} - p_{\ell,M} & -\dfrac{2}{\sqrt{3}}a \\[3mm]
        \dfrac{2}{\sqrt{3}}a & -\dfrac{d}{dr} - q_{\ell,M}
    \end{pmatrix}.
\end{align}
Using the Bogomol'nyi equations \eqref{eq:order-three-Bogomolny-equations}, one finds the factorization
\begin{align}
    H_{\ell,M}^{\perp}
    = \left(\mathcal A_{\ell,M}^{\perp}\right)^\dagger \mathcal A_{\ell,M}^{\perp} + \frac{8}{3}\mathbf 1_2.
\end{align}
Hence,
\begin{align}
    H_{\ell,M}^{\perp}
    \geq \frac{8}{3}\mathbf 1_2.
\end{align}
Combining this with \eqref{eq:order-three-BPS-decoupled-lower-bounds}, we obtain
\begin{align}
    L_\ell^1 \geq \frac{8}{3}\mathbf 1_4,
    \qquad
    \ell\in\mathbb Z.
\end{align}

The similarity relation \eqref{eq:order-three-charged-block-similarity} then implies
\begin{align}
    L_\ell^{-1} \geq \frac{8}{3}\mathbf 1_4, \qquad
    \ell\in\mathbb Z.
\end{align}
Thus, the \(\rho=\pm1\) sectors contain neither negative modes nor zero modes.
Together with the result for the \(\rho=0\) sector, this proves Eq.~\eqref{eq:order-three-BPS-stability-result}.

Therefore, the value \(E_M=4\pi M/9\) obtained from the restricted Bogomol'nyi completion \eqref{eq:order-three-Bogomolny-completion} cannot be lowered at quadratic order by any fluctuation of the full \(SO(3)\) theory.
The quadratic variation vanishes only along the \(2M\) vortex-core position zero modes.

\section{Stability of the order-two vortex family}
\label{sec:order-two-stability}
In this section, we analyze the linear stability of the order-two vortex family constructed in Ref.~\cite{Abe:2026zrk}.
After reviewing the vortex ansatz and the Fourier decomposition of the gauge-fixed Hessian, we study the \(N=1\) vortex numerically and construct an exact negative mode at \(\kappa=\eta=1\).

\subsection{Ansatz and complex basis}
\label{subsec:order-two-ansatz-complex-basis}
We first review the order-two vortex ansatz and introduce a complex basis convenient for the subsequent fluctuation analysis.

We choose \(r_2\) to be the rotation by \(\pi\) about the \(\mathcal Z\)-axis.
With the conventions of Eq.~\eqref{eq:so3-generators}, it is given by
\begin{align}
    r_2 &:= \exp\left(\pi\ii T_3\right)
    =
    \begin{pmatrix}
        -1 & 0 & 0\\
        0 & -1 & 0\\
        0 & 0 & 1
    \end{pmatrix}, &
    r_2^2 &= \boldsymbol{1}_3.
\end{align}
As in the order-three case, the spin-three representation \(\mathcal H_3\) decomposes over \(\mathbb R\) under rotations about the \(\mathcal Z\)-axis into one one-dimensional invariant subspace and three two-dimensional invariant subspaces.
As in the order-three case, one can choose a real orthonormal basis \(\{g_1,\ldots,g_7\}\) of \(\mathcal H_3\) such that
\begin{align}
    X_3 g_1 &= 0, \notag\\
    X_3 g_{2q} &= -q g_{2q+1}, &
    X_3 g_{2q+1} &= q g_{2q},
    \qquad q=1,2,3.
    \label{eq:order-two-X3-real-basis}
\end{align}
An explicit orthonormal basis with this property is
\begin{align}
    g_1 &:= \frac{\sqrt{7}}{2}\mathcal Z \left(-3\mathcal X^2 - 3\mathcal Y^2 + 2\mathcal Z^2\right), \notag\\
    g_2 &:= \frac{\sqrt{42}}{4}\mathcal Y \left(-\mathcal X^2 - \mathcal Y^2 + 4\mathcal Z^2\right), \notag\\
    g_3 &:= \frac{\sqrt{42}}{4}\mathcal X \left(\mathcal X^2 + \mathcal Y^2 - 4\mathcal Z^2\right), \notag\\
    g_4 &:= \frac{\sqrt{105}}{2}\mathcal Z \left(\mathcal X^2 - \mathcal Y^2\right), \notag\\
    g_5 &:= \sqrt{105}\,\mathcal X\mathcal Y\mathcal Z, \notag\\
    g_6 &:= \frac{\sqrt{70}}{4}\mathcal Y \left(-3\mathcal X^2 + \mathcal Y^2\right), \notag\\
    g_7 &:= \frac{\sqrt{70}}{4}\mathcal X \left(\mathcal X^2 - 3\mathcal Y^2\right).
\end{align}
These polynomials are orthonormal with respect to \eqref{eq:H3-inner-product}.
Using
\begin{align}
    X_3 = \mathcal X\frac{\partial}{\partial\mathcal Y} - \mathcal Y\frac{\partial}{\partial\mathcal X},
\end{align}
one directly verifies \eqref{eq:order-two-X3-real-basis}.

The order-two vortex ansatz constructed in Ref.~\cite{Abe:2026zrk}, with \(N\in\mathbb Z_{>0}\), is
\begin{align}
    W &= -\frac{N}{2}\widehat{K}(r)T_3\,d\theta, \label{eq:order-two-gauge-ansatz}\\
    \phi &= \widehat{h}(r) \left[\sin(N\theta)g_4 + \cos(N\theta)g_5\right] \notag\\
    &= \widehat{h}(r) \exp\left(\frac{N\theta}{2}X_3\right)g_5.
    \label{eq:order-two-Higgs-ansatz}
\end{align}
Here \((r,\theta)\) are polar coordinates in the plane transverse to the vortex.
The profile functions satisfy
\begin{align}
    \widehat{K}(0)=0, \qquad
    \widehat{h}(0)=0,
\end{align}
and
\begin{align}
    \widehat{K}(r) \longrightarrow 1, \qquad
    \widehat{h}(r) \longrightarrow 1, \qquad
    (r\longrightarrow\infty).
\end{align}
Consequently,
\begin{align}
    \phi(r,\theta) &\longrightarrow \exp\left(\frac{N\theta}{2}X_3\right) \phi_{\mathrm{vac}}, &
    \phi_{\mathrm{vac}} &:= g_5, \qquad
    (r\longrightarrow\infty),
\end{align}
where \(\phi_{\mathrm{vac}}\) is a representative minimum of the Higgs potential as stated in Eq.~(\ref{eq:higgs-vac}), and the gauge transformation accumulated after one circuit around the vortex is
\begin{align}
    \exp\left(\pi\ii NT_3\right) = r_2^N.
\end{align}
Because \(Q\) in the Higgs potential is homogeneous and \(SO(3)\)-invariant, the Higgs ansatz satisfies
\begin{align}
    Q(\phi) = \widehat{h}^4 Q(g_5) = 0.
\end{align}
Since \(Q\geq0\) and \(Q(g_5)=0\), we have \(\nabla Q(g_5)=0\). Homogeneity and \(SO(3)\) invariance then imply that both \(Q(\phi)\) and \(\nabla Q(\phi)\) vanish on the order-two ansatz. Since the \(\eta\)-dependence of the Higgs potential enters entirely through the coefficient multiplying \(Q(\phi)\), the restricted tension and profile equations are therefore independent of \(\eta\).
Substituting Eqs.~\eqref{eq:order-two-gauge-ansatz} and \eqref{eq:order-two-Higgs-ansatz} into the field equations \eqref{eq:dimensionless-gauge-equation} and \eqref{eq:dimensionless-Higgs-equation}, the nontrivial components reduce to
\begin{align}
    0 &= \widehat{K}'' -\frac{1}{r}\widehat{K}' -4\widehat{h}^2(\widehat{K} - 1), \label{eq:order-two-profile-equation-K}\\
    0 &= \widehat{h}'' + \frac{1}{r}\widehat{h}' - \frac{N^2(1-\widehat{K})^2}{r^2}\widehat{h} - 2\kappa^2(\widehat{h}^2 - 1)\widehat{h}. \label{eq:order-two-profile-equation-h}
\end{align}
The tension restricted to this ansatz is
\begin{align}
    E_N[\widehat{K},\widehat{h}] = \pi\int_0^\infty r\,dr\, \left[
        \frac{N^2(\widehat{K}')^2}{4r^2} + (\widehat{h}')^2 + \frac{N^2(1-\widehat{K})^2\widehat{h}^2}{r^2} + \kappa^2(\widehat{h}^2 - 1)^2
    \right].
    \label{eq:order-two-radial-tension}
\end{align}

At \(\kappa=1\), independently of \(\eta\), the profile equations
\eqref{eq:order-two-profile-equation-K} and \eqref{eq:order-two-profile-equation-h}, together with the restricted tension \eqref{eq:order-two-radial-tension}, coincide with those of the critically coupled Abelian-Higgs \(N\)-vortex in the present normalization \cite{Bogomolny:1975de,Jaffe:1980mj,Abe:2026zrk}.
The restricted tension admits the Bogomol'nyi completion
\begin{align}
    E_N &= \pi\int_0^\infty r\,dr\, \left[
        \left\{\frac{N\widehat{K}'}{2r} - (1-\widehat{h}^2)\right\}^2 + \left\{\widehat{h}' - \frac{N(1-\widehat{K})}{r}\widehat{h}\right\}^2
    \right]
    + \pi N \notag\\
    &\geq \pi N.
    \label{eq:order-two-Bogomolny-completion}
\end{align}
The regular finite-tension solution of Eqs.~\eqref{eq:order-two-profile-equation-K} and \eqref{eq:order-two-profile-equation-h} satisfies the Bogomol'nyi equations
\begin{align}
    \widehat{h}' &= \frac{N(1-\widehat{K})}{r}\widehat{h}, &
    \frac{N\widehat{K}'}{r} &= 2(1-\widehat{h}^2),
    \label{eq:order-two-Bogomolny-equations}
\end{align}
and therefore saturates the bound,
\begin{align}
    E_N = \pi N.
\end{align}

The inequality above establishes a Bogomol'nyi bound only for field configurations of the form \eqref{eq:order-two-gauge-ansatz} and \eqref{eq:order-two-Higgs-ansatz}.
It does not imply the same lower bound for the full \(SO(3)\) theory, because the full theory also allows gauge and Higgs fluctuations
outside this ansatz.

For the stability analysis, we complexify the gauge and Higgs fields.
For the gauge field, we introduce
\begin{align}
    \widehat{T}^0 &:= T_3, &
    \widehat{T}^c &:= -T_1, &
    \widehat{T}^s &:= -T_2, \notag\\
    \widehat{T}^+ &:= \frac{\widehat{T}^c + \ii \widehat{T}^s}{\sqrt{2}}, &
    \widehat{T}^- &:= \frac{\widehat{T}^c - \ii \widehat{T}^s}{\sqrt{2}}.
\end{align}
The corresponding generators acting on the Higgs field are
\begin{align}
    \widehat{X}^0 &:= X_3, &
    \widehat{X}^c &:= -X_1, &
    \widehat{X}^s &:= -X_2, \notag\\
    \widehat{X}^+ &:= \frac{\widehat{X}^c + \ii \widehat{X}^s}{\sqrt{2}}, &
    \widehat{X}^- &:= \frac{\widehat{X}^c - \ii \widehat{X}^s}{\sqrt{2}}.
\end{align}
With these conventions,
\begin{align}
    [\widehat{X}^0,\widehat{X}^+] &= \ii \widehat{X}^+, &
    [\widehat{X}^0,\widehat{X}^-] &= -\ii \widehat{X}^-.
\end{align}

For the Higgs field, we define the complex basis
\begin{align}
    \widehat{u}_0 &:= g_1, \notag\\
    \widehat{u}_q &:= \frac{g_{2q}+\ii g_{2q+1}}{\sqrt{2}}, &
    \widehat{u}_{-q} &:= \frac{g_{2q}-\ii g_{2q+1}}{\sqrt{2}},
    \qquad q=1, 2, 3.
\end{align}
It satisfies
\begin{align}
    \widehat{X}^0 \widehat{u}_q = \ii q\,\widehat{u}_q, \qquad
    q=0, \pm1, \pm2, \pm3,
\end{align}
and, with the complex-bilinear extension of the inner product,
\begin{align}
    \langle \widehat{u}_q, \widehat{u}_{q'}\rangle_\ip = \delta_{q+q',0}.
\end{align}

In the complex basis, the vortex background \eqref{eq:order-two-gauge-ansatz} and \eqref{eq:order-two-Higgs-ansatz} becomes
\begin{align}
    W &= -\frac{N}{2}\widehat{K}(r)\widehat{T}^0\,d\theta, \\
    \phi &= -\frac{\ii \widehat{h}(r)}{\sqrt{2}} \left[e^{\ii N\theta}\widehat{u}_2 - e^{-\ii N\theta}\widehat{u}_{-2}\right].
\end{align}
The complexified fluctuations are expanded as
\begin{align}
    B_i &= \widehat{B}_i^0\widehat{T}^0+\widehat{B}_i^+\widehat{T}^++\widehat{B}_i^-\widehat{T}^-, \notag\\
    \xi &= \sum_{q=-3}^{3}\widehat{\xi}_q\widehat{u}_q.
\end{align}
We also complexify the spatial vector components by defining
\begin{align}
    \widehat{B}_+^q &:= \frac{\widehat{B}_1^q-\ii \widehat{B}_2^q}{\sqrt{2}}, &
    \widehat{B}_-^q &:= \frac{\widehat{B}_1^q+\ii \widehat{B}_2^q}{\sqrt{2}}, \qquad
    q=0, \pm1.
\end{align}

We denote the complex-linear extension of the gauge-fixed Hessian \(\widetilde L\) by \(\widetilde{\widetilde L}\).
The complex bases introduced above will be used to decompose \(\widetilde{\widetilde L}\) into angular Fourier blocks.

\subsection{Fourier blocks and symmetry modes}
\label{subsec:order-two-fourier-blocks}
We use the complexified fluctuation space \(\widetilde{\mathcal X}\) and the radial Hilbert space \(\mathscr H_r=L^2((0,\infty),r\,dr)\) introduced in Sec.~\ref{subsec:order-three-fourier-blocks}.
The thirteen complex components of a fluctuation are
\begin{align}
    \bigl(
        \widehat{B}_+^0,\,
        \widehat{B}_-^0,\,
        \widehat{B}_+^+,\,
        \widehat{B}_-^+,\,
        \widehat{B}_+^-,\,
        \widehat{B}_-^-,\,
        \widehat{\xi}_0,\,
        \widehat{\xi}_{+1},\,
        \widehat{\xi}_{-1},\,
        \widehat{\xi}_{+2},\,
        \widehat{\xi}_{-2},\,
        \widehat{\xi}_{+3},\,
        \widehat{\xi}_{-3}
    \bigr).
\end{align}
For the stability analysis, we organize the Fourier modes into four
classes. Since the Higgs background contains the components
\(e^{\pm \ii N\theta}\widehat u_{\pm2}\), the corresponding Higgs
fluctuations acquire Fourier-index shifts by \(\pm N\). The
\(\rho=0\) sector contains the neutral gauge components
\(\widehat B^0\) together with \(\widehat\xi_{\pm2}\), while the gauge
components \(\widehat B^\pm\) couple to the Higgs fluctuations obtained
by acting with \(\widehat X^\pm\) on the background. Thus, the
\(\rho=1\) sector contains \(\widehat\xi_{+3}\) and
\(\widehat\xi_{-1}\), whereas the \(\rho=-1\) sector contains
\(\widehat\xi_{+1}\) and \(\widehat\xi_{-3}\). The component
\(\widehat\xi_0\) forms a separate neutral sector, denoted by
\(\rho=\mathrm n\). In addition, the spatially complexified gauge
components \(\widehat B_\pm\) carry angular helicity \(\pm1\), leading
to the shifts \(\ell\mp1\).
Thus, for each \(\ell\in\mathbb Z\), we decompose the fluctuation space as
\begin{align}
    \widetilde{\mathcal X} = \bigoplus_{\ell\in\mathbb Z}
    \left(
        \widehat{\mathcal F}_\ell^{\mathrm n} \oplus
        \widehat{\mathcal F}_\ell^0 \oplus
        \widehat{\mathcal F}_\ell^1 \oplus
        \widehat{\mathcal F}_\ell^{-1}
    \right).
\end{align}
They are defined by
\begin{align}
    \widehat{\mathcal F}_\ell^{\mathrm n} &:=
    \underbrace{e^{\ii\ell\theta}\mathscr H_r}_{\widehat{\xi}_0}, \\[1ex]
    \widehat{\mathcal F}_\ell^0 &:=
    \underbrace{e^{\ii(\ell-1)\theta}\mathscr H_r}_{\widehat{B}_+^0} \oplus
    \underbrace{e^{\ii(\ell+1)\theta}\mathscr H_r}_{\widehat{B}_-^0} \oplus
    \underbrace{e^{\ii(\ell+N)\theta}\mathscr H_r}_{\widehat{\xi}_{+2}} \oplus
    \underbrace{e^{\ii(\ell-N)\theta}\mathscr H_r}_{\widehat{\xi}_{-2}}, \\[1ex]
    \widehat{\mathcal F}_\ell^1 &:=
    \underbrace{e^{\ii(\ell-1)\theta}\mathscr H_r}_{\widehat{B}_+^+} \oplus
    \underbrace{e^{\ii(\ell+1)\theta}\mathscr H_r}_{\widehat{B}_-^+} \oplus
    \underbrace{-\ii e^{\ii(\ell+N)\theta}\mathscr H_r}_{\widehat{\xi}_{+3}} \oplus
    \underbrace{-\ii e^{\ii(\ell-N)\theta}\mathscr H_r}_{\widehat{\xi}_{-1}},
    \label{eq:order-two-Fourier-space-rho1} \\[1ex]
    \widehat{\mathcal F}_\ell^{-1} &:=
    \underbrace{e^{\ii(\ell-1)\theta}\mathscr H_r}_{\widehat{B}_+^-} \oplus
    \underbrace{e^{\ii(\ell+1)\theta}\mathscr H_r}_{\widehat{B}_-^-} \oplus
    \underbrace{\ii e^{\ii(\ell+N)\theta}\mathscr H_r}_{\widehat{\xi}_{+1}} \oplus
    \underbrace{\ii e^{\ii(\ell-N)\theta}\mathscr H_r}_{\widehat{\xi}_{-3}}.
    \label{eq:order-two-Fourier-space-rho-minus1}
\end{align}
Here, omitted components are understood to vanish.
The constant phases in Eqs.~\eqref{eq:order-two-Fourier-space-rho1} and \eqref{eq:order-two-Fourier-space-rho-minus1} are chosen so that the corresponding radial operators have real coefficients.

Direct substitution into the complexified gauge-fixed Hessian gives
\begin{align}
    \widetilde{\widetilde L}\,
    \widehat{\mathcal F}_\ell^\rho &\subset \widehat{\mathcal F}_\ell^\rho, \qquad
    \rho = \mathrm n, 0,\pm1.
\end{align}
Thus, modes with different labels \((\rho,\ell)\) do not mix.

After factoring out the prescribed angular dependences, we introduce the radial coefficient vectors
\begin{align}
    \widehat{u}_\ell^{\mathrm n}(r) &:=
    \widehat{\chi}_0(r), \notag\\
    \widehat{u}_\ell^0(r) &:=
    \begin{pmatrix}
        \widehat{S}_+^0(r)\\
        \widehat{S}_-^0(r)\\
        \widehat{\chi}_{+2}(r)\\
        \widehat{\chi}_{-2}(r)
    \end{pmatrix}, &
    \widehat{u}_\ell^1(r) &:=
    \begin{pmatrix}
        \widehat{S}_+^+(r)\\
        \widehat{S}_-^+(r)\\
        \widehat{\chi}_{+3}(r)\\
        \widehat{\chi}_{-1}(r)
    \end{pmatrix}, &
    \widehat{u}_\ell^{-1}(r) &:=
    \begin{pmatrix}
        \widehat{S}_+^-(r)\\
        \widehat{S}_-^-(r)\\
        \widehat{\chi}_{+1}(r)\\
        \widehat{\chi}_{-3}(r)
    \end{pmatrix}.
\end{align}
The corresponding angular factors are
\begin{align}
    \widehat{U}_\ell^{\mathrm n} &:= e^{\ii\ell\theta}, \notag\\
    \widehat{U}_\ell^0 &:= \operatorname{diag}\left(
        e^{\ii(\ell-1)\theta},
        e^{\ii(\ell+1)\theta},
        e^{\ii(\ell+N)\theta},
        e^{\ii(\ell-N)\theta}
    \right), \notag\\
    \widehat{U}_\ell^1 &:= \operatorname{diag}\left(
        e^{\ii(\ell-1)\theta},
        e^{\ii(\ell+1)\theta},
        -\ii e^{\ii(\ell+N)\theta},
        -\ii e^{\ii(\ell-N)\theta}
    \right), \notag\\
    \widehat{U}_\ell^{-1} &:= \operatorname{diag}\left(
        e^{\ii(\ell-1)\theta},
        e^{\ii(\ell+1)\theta},
        \ii e^{\ii(\ell+N)\theta},
        \ii e^{\ii(\ell-N)\theta}
    \right).
\end{align}
The radial operators are defined by
\begin{align}
    \widetilde{\widetilde L} \left(\widehat{U}_\ell^\rho \widehat{u}_\ell^\rho\right)
    = \widehat{U}_\ell^\rho \left(\widehat{L}_\ell^\rho \widehat{u}_\ell^\rho\right), \qquad
    \rho = \mathrm n,0,\pm1.
    \label{eq:order-two-radial-block-definition}
\end{align}
They take the Schrödinger-type form
\begin{align}
    \widehat{L}_\ell^{\mathrm n} &= -\Delta_r + \widehat{V}_\ell^{\mathrm n}(r), \notag\\
    \widehat{L}_\ell^\rho &= -\Delta_r\,\mathbf 1_4 + \widehat{V}_\ell^\rho(r), \qquad
    \rho = 0,\pm1,
\end{align}
where
\begin{align}
    \Delta_r := \frac{d^2}{dr^2} + \frac{1}{r}\frac{d}{dr}.
\end{align}
The functions \(\widehat{V}_\ell^{\mathrm n}\) and the \(4\times4\) matrices \(\widehat{V}_\ell^\rho\) are real, and the latter are symmetric.
Their explicit expressions are listed in Appendix~\ref{app:order-two-radial-blocks}.

As in the order-three family, complex conjugation commutes with the complexified gauge-fixed Hessian,
\begin{align}
    \widetilde{\widetilde L}\,\overline{\Psi}
    = \overline{\widetilde{\widetilde L}\Psi},
\end{align}
and maps the Fourier subspaces according to
\begin{align}
    \overline{\widehat{\mathcal F}_\ell^{\mathrm n}}
    &= \widehat{\mathcal F}_{-\ell}^{\mathrm n}, &
    \overline{\widehat{\mathcal F}_\ell^0}
    &= \widehat{\mathcal F}_{-\ell}^0, &
    \overline{\widehat{\mathcal F}_\ell^1}
    &= \widehat{\mathcal F}_{-\ell}^{-1}.
\end{align}
Consequently, by the same argument as in Sec.~\ref{subsec:order-three-fourier-blocks},
\begin{align}
    \widehat{L}_\ell^{\mathrm n} &= \widehat{L}_{-\ell}^{\mathrm n}, \notag\\
    P\widehat{L}_\ell^0P^{-1} &= \widehat{L}_{-\ell}^0, \notag\\
    P\widehat{L}_\ell^1P^{-1} &= \widehat{L}_{-\ell}^{-1},
    \label{eq:order-two-block-similarity-relations}
\end{align}
where \(P\) is defined in Eq.~\eqref{eq:order-three-charged-block-similarity}.
These relations can also be verified directly from the explicit radial operators given in Appendix~\ref{app:order-two-radial-blocks}.
It is therefore sufficient to analyze
\begin{align}
    \rho=\mathrm n,0, \quad \ell\geq0,
    \qquad\text{and}\qquad
    \rho=1, \quad \ell\in\mathbb Z.
\end{align}

As in the order-three family, the zero modes of the gauge-fixed Hessian associated with infinitesimal translations of the vortex center are
\begin{align}
    \widehat{Z}_j &:= \begin{pmatrix}
        G_{ji}\\
        D_j\phi
    \end{pmatrix},
    \qquad j=1,2.
\end{align}
They satisfy the background gauge condition and
\begin{align}
    \widetilde L \widehat{Z}_j = 0,
    \qquad j=1,2.
\end{align}
The combinations
\begin{align}
    \widehat Z^{(+)} &:= -\ii\left(\widehat{Z}_1+\ii \widehat{Z}_2\right), &
    \widehat Z^{(-)} &:= \ii\left(\widehat{Z}_1-\ii \widehat{Z}_2\right)
\end{align}
are complex translational zero modes belonging to \(\widehat{\mathcal F}_1^0\) and \(\widehat{\mathcal F}_{-1}^0\), respectively.
Substituting the order-two vortex background into the translational zero modes, we obtain
\begin{align}
    \widehat Z^{(+)} &= \widehat{U}_1^0\, \frac{1}{\sqrt{2}}
    \begin{pmatrix}
        \dfrac{N\widehat{K}'}{r}\\[2mm]
        0\\[1mm]
        -\widehat{h}'+\dfrac{N(1-\widehat{K})}{r}\widehat{h}\\[2mm]
        \widehat{h}'+\dfrac{N(1-\widehat{K})}{r}\widehat{h}
    \end{pmatrix}
    \in \widehat{\mathcal F}_1^0, \\[3mm]
    \widehat Z^{(-)} &= \widehat{U}_{-1}^0\, \frac{1}{\sqrt{2}}
    \begin{pmatrix}
        0\\[1mm]
        \dfrac{N\widehat{K}'}{r}\\[2mm]
        \widehat{h}'+\dfrac{N(1-\widehat{K})}{r}\widehat{h}\\[2mm]
        -\widehat{h}'+\dfrac{N(1-\widehat{K})}{r}\widehat{h}
    \end{pmatrix}
    \in \widehat{\mathcal F}_{-1}^0.
\end{align}
They satisfy
\begin{align}
    \widetilde{\widetilde L}\widehat Z^{(\pm)} = 0, \qquad
    \widehat Z^{(-)} = \overline{\widehat Z^{(+)}}.
\end{align}

Finally, as in the order-three family, the infinitesimal rotational deformation is a gauge zero mode generated by the constant parameter \(-N\widehat{T}^0/2\).
It is lifted by the background gauge fixing and therefore does not appear as an additional zero mode of \(\widetilde{\widetilde L}\).

\subsection{Numerical results}
\label{subsec:order-two-numerical-result}
We numerically solved the radial equations \eqref{eq:order-two-profile-equation-K} and \eqref{eq:order-two-profile-equation-h} and substituted the resulting profiles into the radial operators \(\widehat{L}_\ell^\rho\).
Although the background profiles are independent of \(\eta\) because
\(Q(\phi)=\nabla Q(\phi)=0\), the fluctuation operators retain an
\(\eta\)-dependence through the generally nonvanishing Hessian
\(H_Q(\phi)\).

Let \(\widehat{\lambda}_{\rho,\ell}^{(N)}(\kappa,\eta)\) denote the lowest eigenvalue of \(\widehat{L}_\ell^\rho\), excluding the translational zero mode, and define
\begin{align}
    \widehat{\lambda}_N(\kappa,\eta) :=
    \min_{\substack{\rho\in\{\mathrm n,0,\pm1\}\\ \ell\in\mathbb Z}}
    \widehat{\lambda}_{\rho,\ell}^{(N)}(\kappa,\eta).
\end{align}
The similarity relations \eqref{eq:order-two-block-similarity-relations} allow us to take \(\rho=\mathrm n,0\) with \(\ell\geq0\) and \(\rho=1\) with \(\ell\in\mathbb Z\) as independent representatives.

\begin{figure}[t]
    \centering
    \includegraphics[width=0.72\textwidth]{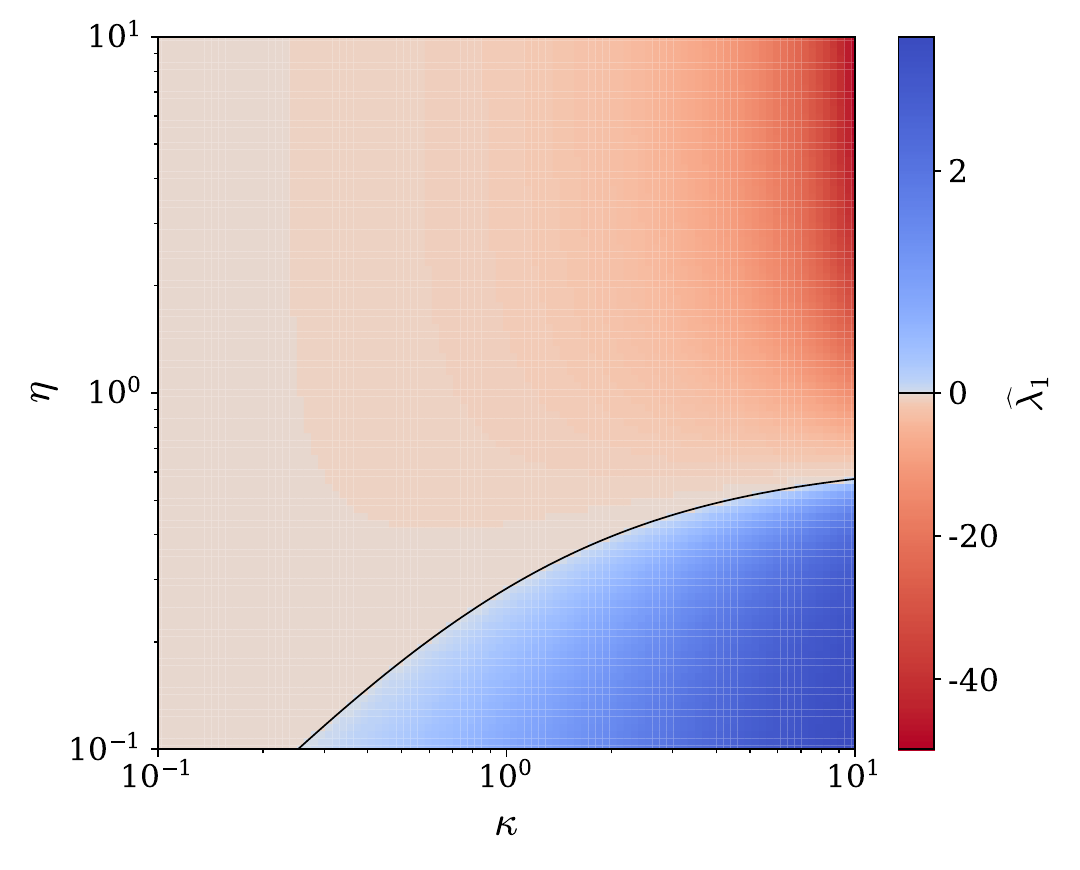}
    \caption{
        Minimum eigenvalue \(\widehat{\lambda}_1(\kappa,\eta)\) of the gauge-fixed Hessian for the \(N=1\) order-two vortex.
        The minimum is attained throughout the displayed region in the \((\rho,\ell)=(1,1)\) block, or equivalently in \((\rho,\ell)=(-1,-1)\).
        The black curve represents \(\widehat{\lambda}_1(\kappa,\eta)=0\).
    }
    \label{fig:order-two-N1-minimum-eigenvalue}
\end{figure}

Figure~\ref{fig:order-two-N1-minimum-eigenvalue} shows \(\widehat{\lambda}_1(\kappa,\eta)\) for \(N=1\).
At every parameter point in the numerical scan, the minimum is attained in the \((\rho,\ell)=(1,1)\) block, or equivalently in the conjugate block \((\rho,\ell)=(-1,-1)\).
This behavior is naturally understood from the fact that, for \(N=1\) and \(\ell=1\), the \(\widehat B_+^+\) and \(\widehat\xi_{-1}\) components carry no additional angular dependence and are coupled by an off-diagonal term in the Hessian as seen later.
The black curve represents \(\widehat{\lambda}_1(\kappa,\eta)=0\).
The \(N=1\) order-two vortex therefore possesses a negative mode over a broad region of the parameter space.

Along the entire line \(\kappa=1\), independently of \(\eta\), the order-two background satisfies the Bogomol'nyi equations \eqref{eq:order-two-Bogomolny-equations} and saturates the bound \eqref{eq:order-two-Bogomolny-completion}.
Nevertheless, its stability depends on \(\eta\):
for sufficiently small \(\eta\) in the displayed region, no negative mode is detected, while at \((\kappa,\eta)=(1,1)\) the next subsection proves analytically that the \((\rho,\ell)=(1,1)\) block has an exact eigenvalue \(-1\).
Thus, saturation of the Bogomol'nyi bound within the restricted vortex ansatz does not imply stability against general fluctuations of the full \(SO(3)\) theory.

\subsection{Instability at \texorpdfstring{\(\kappa=\eta=1\)}{kappa=eta=1}}
\label{subsec:order-two-exact-negative-mode}
The point \((\kappa,\eta)=(1,1)\) sharply distinguishes the two vortex families.
For the order-three family, the gauge-fixed Hessian is non-negative, as shown in Sec.~\ref{subsec:order-three-analytical-stability}.
By contrast, the \(N=1\) order-two vortex has an exact eigenvalue \(-1\) in the \((\rho,\ell)=(1,1)\) Fourier block and in the symmetry-related \((\rho,\ell)=(-1,-1)\) block.

We first consider the \((\rho,\ell)=(1,1)\) block.
With the radial components ordered as \(\left(\widehat{S}_+^+,\,\widehat{S}_-^+,\,\widehat{\chi}_{+3},\,\widehat{\chi}_{-1}\right)^{\mathsf T}\), the following mode has eigenvalue \(-1\):
\begin{align}
    \omega(r) &:= f(r) \begin{pmatrix} 1\\ 0\\ 0\\ 1 \end{pmatrix}, \label{eq:order-two-N1-minus-one-radial-mode} \\
    f(r) &:= \sqrt{\frac{\widehat{h}(r)}{r}}\, \exp\left[-\sqrt{5}\int_0^r \widehat{h}(s)\,ds\right]. \label{eq:order-two-N1-minus-one-profile}
\end{align}

To verify that this mode has eigenvalue \(-1\), recall that, at \(\kappa=1\), the order-two background satisfies the Bogomol'nyi equations \eqref{eq:order-two-Bogomolny-equations}.
For \(N=1\), they read
\begin{align}
    \widehat{h}' &= \frac{1-\widehat{K}}{r}\widehat{h}, &
    \frac{\widehat{K}'}{r} &= 2(1-\widehat{h}^2).
    \label{eq:order-two-N1-Bogomolny-equations}
\end{align}
The explicit matrix form of \(\widehat{L}_\ell^1\) is given in Appendix~\ref{app:order-two-radial-blocks}.
Setting \((\kappa,\eta)=(1,1)\), \(N=1\), and \(\ell=1\), and using Eq.~\eqref{eq:order-two-N1-Bogomolny-equations}, we obtain
\begin{align}
    \widehat{L}_1^1 &= -\Delta_r\,\mathbf 1_4 + \widehat{V}_1^1, \notag\\
    \widehat{V}_1^1 &=
    \begin{pmatrix}
        \dfrac{\widehat{K}^2}{4r^2} + 6\widehat{h}^2 - 2 & 0 & 0 & -2\sqrt{5}\,\widehat{h}' \\[3mm]
        0 & \dfrac{(4-\widehat{K})^2}{4r^2} + 2\widehat{h}^2 + 2 & -2\sqrt{3}\,\widehat{h}' & 0 \\[3mm]
        0 & -2\sqrt{3}\,\widehat{h}' & \dfrac{(4-3\widehat{K})^2}{4r^2} + 6\widehat{h}^2 - 2 & 0 \\[3mm]
        -2\sqrt{5}\,\widehat{h}' & 0 & 0 & \dfrac{\widehat{K}^2}{4r^2} + 6\widehat{h}^2 - 2
    \end{pmatrix}.
\end{align}
Using the definition \eqref{eq:order-two-N1-minus-one-profile} and the Bogomol'nyi equations \eqref{eq:order-two-N1-Bogomolny-equations}, we obtain
\begin{align}
    f' = -\left(\frac{\widehat{K}}{2r}+\sqrt{5}\,\widehat{h}\right)f.
\end{align}
Differentiating this relation and using Eq.~\eqref{eq:order-two-N1-Bogomolny-equations}, one directly verifies
\begin{align}
    \widehat{L}_1^1\omega = -\omega.
\end{align}
Thus, the \(N=1\) order-two vortex has an exact negative mode at \((\kappa,\eta)=(1,1)\).

The \((\rho,\ell)=(1,1)\) block contains the fluctuation components
\(\widehat{B}_+^+\), \(\widehat{B}_-^+\), \(\widehat{\xi}_{+3}\), and
\(\widehat{\xi}_{-1}\), which lie outside the order-two vortex ansatz.
The exact negative mode has nonzero \(\widehat{B}_+^+\) and
\(\widehat{\xi}_{-1}\) components. Since \(\widehat{B}_+^+\) probes the
generator \(\widehat{T}^+\), which does not commute with the background
generator \(\widehat{T}^0\), this mode directly reflects the non-Abelian
structure of the full \(SO(3)\) theory and its role in vortex stability.

\section{Summary and discussion}
\label{sec:summary}

We have studied the linear energetic stability of the order-three and order-two vortex families constructed in Ref.~\cite{Abe:2026zrk} in an \(SO(3)\) gauge--Higgs theory with residual gauge group \(A_4\).
After fixing the gauge and expanding the perturbations in Fourier modes of the polar angle, we have reduced the two-dimensional fluctuation problem to one-dimensional radial eigenvalue problems.
We have analyzed these problems analytically at special parameter values and numerically over a broad region of the \((\kappa,\eta)\) parameter space.

For the order-three family, no negative mode has been found for \(M=1\) in the scanned parameter range, whereas an unstable region has been found for \(M=2\).
At \((\kappa,\eta)=(1,1)\), we have proved for every \(M\geq1\) that the full gauge-fixed Hessian is non-negative and has precisely \(2M\) physical zero modes.
These zero modes correspond to infinitesimal variations of the positions of the \(M\) vortex cores in the transverse plane.
Thus, at this parameter point, no physical perturbation of the full \(SO(3)\) theory lowers the tension at quadratic order.

The \(N=1\) order-two vortex behaves differently.
Its lowest eigenvalue is negative over a broad region of the scanned parameter space, and throughout the scan the minimum is attained in the \((\rho,\ell)=(1,1)\) block, or equivalently in the symmetry-related \((\rho,\ell)=(-1,-1)\) block.
At \((\kappa,\eta)=(1,1)\), we have constructed an exact eigenmode with eigenvalue \(-1\).
This sharp contrast with the order-three family shows that the reduced Bogomol'nyi structure alone does not determine stability in the full \(SO(3)\) theory.

The \(M=2\) order-three vortex also illustrates the distinction between local spectral stability and energetic binding.
The zero contours of \(\lambda_{0,2}^{(2)}\) and \(\Delta_{\mathrm{bin}}=E_2-2E_1\) do not coincide, and we have found a region with
\begin{align}
    \Delta_{\mathrm{bin}} < 0, \qquad
    \lambda_{0,2}^{(2)} < 0.
\end{align}
This result suggests a possible molecular bound state at finite vortex separation.
Establishing whether such a state exists requires a nonlinear analysis and is left for future work.

Any continuous nonlinear evolution that preserves the asymptotic boundary conditions must preserve the total topological charge.
The relation between this global topological structure and the local stability properties found here deserves further investigation.

It is also important that topological triviality does not preclude local stability.
For example, the order-three configuration with \(M=6\) lies in the same topological sector as the vacuum \cite{Abe:2026zrk}, yet it is locally stable in a finite region of parameter space.\footnote{Additional numerical scans, not shown here, find the gauge-fixed Hessian to be positive apart from the translational zero modes in such a region.}
Its decay to the vacuum would then require a finite-amplitude deformation beyond the linearized regime.

The choice of \(A_4\) as the residual gauge group also connects the present model to broader developments in particle physics.
The group \(A_4\) has been widely used as a discrete flavor symmetry in models of lepton masses and mixing \cite{Ma:2001dn,Altarelli:2005yp,Altarelli:2010gt,Ishimori:2010au}, and its realization as a residual subgroup of \(SO(3)\) has been studied in several gauge--Higgs constructions \cite{Ovrut:1977cn,Etesi:1996urw,King:2018fke}.
Reference.~\cite{Abe:2026zrk} further related the smooth finite-tension vortices of the present theory to Gukov--Witten surface operators \cite{Gukov:2008sn} associated with the non-invertible one-form symmetry of the infrared \(A_4\) gauge theory.
How the stability of the smooth vortex configurations is reflected in this infrared description 
(see Refs.~\cite{Heidenreich:2021xpr,Hidaka:2024kfx}) 
remains an interesting open question.

Although the present analysis concerns static straight vortices, 
the stability analysis of junctions \cite{Kobayashi:2008pk} and 
non-Abelian knots \cite{Brekke:1997jj,Annala:2022bdd,Rajamaki:2023ymv,Kobayashi:2024aip} 
remains a future problem.
The results may also have implications for non-Abelian cosmic strings.
If an \(SO(3)\to A_4\) gauge--Higgs phase were realized in a cosmological setting, the vortices studied here would provide examples of non-Abelian cosmic strings \cite{McGraw:1997nx,Abe:2026zrk}.
When two non-Abelian cosmic strings with non-commutative $\pi_1$ elements collide, they form a
junction, 
in contrast to 
Abelian strings 
resulting in a reconnection 
in collision, 
see e.~g. Ref.~\cite{Eto:2006db}.
Therefore, we expect a non-Abelian cosmic string network 
\cite{McGraw:1997nx}.
The possible molecular bound states and the negative modes found here may provide microscopic mechanisms for string rearrangement or fragmentation.
Such processes could affect the relative populations of string species and the formation of junctions in a non-Abelian string network.
A detailed understanding of these effects will require time-dependent field-theory simulations of the string dynamics.
Collision of global non-Abelian strings was studied 
in Ref.~\cite{Kobayashi:2008pk}.

Gauge--Higgs theories in which \(SO(3)\) is Higgsed to other finite non-Abelian subgroups have also been studied \cite{Ovrut:1977cn,Etesi:1996urw,King:2018fke}.
After constructing the corresponding vortex solutions, one could analyze their stability by following the same strategy used in the present work:
fixing the gauge and decomposing the fluctuations into Fourier modes, as in the approach of Gustafson and Sigal \cite{GustafsonSigal2000}.
Such comparisons may clarify which stability properties are specific to the \(A_4\) model and which are generic to gauge theories with finite non-Abelian residual symmetry.

%%%%%%%%%%%%%%%%%%%%%%%
\section*{Acknowledgments}
\noindent
This work is supported by JST Grant No. JPMJPF2221 (YA),
by JST SPRING, Japan Grant No. JPMJSP2123 (KM),
by JSPS KAKENHI Grant No. JP22K03601 (TH),
JP22H01221 and JP23K22492 (MN), 
JP25K17394 (RY), and by 
the WPI program ``Sustainability with Knotted
 Chiral Meta Matter (WPI-SKCM$^2$)'' at Hiroshima University (MN).

\appendix
\section{Explicit radial block operators for the order-three family}
\label{app:order-three-radial-blocks}
We list here the radial block operators defined in Eq.~\eqref{eq:order-three-radial-block-definition}.
Introduce the radial coefficient vectors
\begin{align}
    u_\ell^0 &:=
    \begin{pmatrix}
        S_+^0&
        S_-^0&
        \chi_0&
        \chi_{+3}&
        \chi_{-3}
    \end{pmatrix}^{\mathsf T},
    \notag\\
    u_\ell^1
    &:=
    \begin{pmatrix}
        S_+^+&
        S_-^+&
        \chi_{+1}&
        \chi_{-2}
    \end{pmatrix}^{\mathsf T},
    \notag\\
    u_\ell^{-1}
    &:=
    \begin{pmatrix}
        S_+^-&
        S_-^-&
        \chi_{+2}&
        \chi_{-1}
    \end{pmatrix}^{\mathsf T}.
\end{align}
The three operators take the form
\begin{align}
    L_\ell^\rho = -\Delta_r\,\mathbf 1 + V_\ell^\rho(r), \qquad
    \Delta_r := \frac{d^2}{dr^2} + \frac{1}{r}\frac{d}{dr},
\end{align}
where \(V_\ell^\rho\) are real symmetric matrices.

\subsection{\texorpdfstring{\(\rho=0\)}{rho=0} block}
In the component order specified by \(u_\ell^0\), the matrix \(V_\ell^0\) is
\begin{align}
    V_\ell^0 =
    \begin{pmatrix}
        V_{11}^{(0)} & 0            & 0            & V_{14}^{(0)} & V_{15}^{(0)} \\
        0            & V_{22}^{(0)} & 0            & V_{24}^{(0)} & V_{25}^{(0)} \\
        0            & 0            & V_{33}^{(0)} & V_{34}^{(0)} & V_{35}^{(0)} \\
        V_{14}^{(0)} & V_{24}^{(0)} & V_{34}^{(0)} & V_{44}^{(0)} & V_{45}^{(0)} \\
        V_{15}^{(0)} & V_{25}^{(0)} & V_{35}^{(0)} & V_{45}^{(0)} & V_{55}^{(0)}
    \end{pmatrix},
\end{align}
where
\begin{align}
    V_{11}^{(0)} &= \frac{(\ell-1)^2}{r^2} + 4a^2, \notag\\[1mm]
    V_{22}^{(0)} &= \frac{(\ell+1)^2}{r^2} + 4a^2, \notag\\[1mm]
    V_{33}^{(0)} &= \frac{\ell^2}{r^2} - 2\kappa^2 + \frac{8\kappa^2}{9}\left(1 - \frac{1}{\eta^2}\right)a^2 + \frac{2\kappa^2}{3}\left(5 + \frac{4}{\eta^2}\right)b^2, \notag\\[1mm]
    V_{44}^{(0)} &= \frac{\bigl[\ell+M(1-K)\bigr]^2}{r^2} + 2a^2 - 2\kappa^2 + \frac{4\kappa^2}{9} \left(4 + \frac{5}{\eta^2}\right)a^2 + \frac{10\kappa^2}{9}\left(1 - \frac{1}{\eta^2}\right)b^2, \notag\\[1mm]
    V_{55}^{(0)} &= \frac{\bigl[\ell - M(1-K)\bigr]^2}{r^2} + 2a^2 - 2\kappa^2 + \frac{4\kappa^2}{9}\left(4 + \frac{5}{\eta^2}\right)a^2 + \frac{10\kappa^2}{9}\left(1 - \frac{1}{\eta^2}\right)b^2, \notag\\[1mm]
    V_{14}^{(0)} = V_{25}^{(0)} &= -2\left[a' - \frac{M(1-K)}{r}a\right], \notag\\
    V_{15}^{(0)} = V_{24}^{(0)} &= -2\left[a' + \frac{M(1-K)}{r}a\right],\notag\\
    V_{34}^{(0)} &= \frac{4\sqrt{10}\kappa^2}{9}\left(1 - \frac{1}{\eta^2}\right)ab, \notag\\
    V_{35}^{(0)} &= -\frac{4\sqrt{10}\kappa^2}{9} \left(1 - \frac{1}{\eta^2}\right)ab, \notag\\
    V_{45}^{(0)} &= 2a^2 - \frac{2\kappa^2}{9}\left(4 + \frac{5}{\eta^2}\right)a^2.
\end{align}

\subsection{\texorpdfstring{\(\rho=1\)}{rho=1} block}
In the component order specified by \(u_\ell^1\), the matrix \(V_\ell^1\) is
\begin{align}
    V_\ell^1 =
    \begin{pmatrix}
        V_{11}^{(1)} & 0            & V_{13}^{(1)} & V_{14}^{(1)} \\
        0            & V_{22}^{(1)} & V_{23}^{(1)} & V_{24}^{(1)} \\
        V_{13}^{(1)} & V_{23}^{(1)} & V_{33}^{(1)} & V_{34}^{(1)} \\
        V_{14}^{(1)} & V_{24}^{(1)} & V_{34}^{(1)} & V_{44}^{(1)}
    \end{pmatrix},
\end{align}
where
\begin{align}
    V_{11}^{(1)} &= \frac{\left(\ell-1-\dfrac{MK}{3}\right)^2}{r^2} - \frac{2MK'}{3r} + \frac{2}{3}a^2 + \frac{10}{3}b^2, \notag\\[1mm]
    V_{22}^{(1)} &= \frac{\left(\ell+1-\dfrac{MK}{3}\right)^2}{r^2} + \frac{2MK'}{3r} + \frac{2}{3}a^2 + \frac{10}{3}b^2, \notag\\[1mm]
    V_{33}^{(1)} &= \frac{\left(\ell-\dfrac{MK}{3}\right)^2}{r^2} + \frac{10}{3}b^2 - 2\kappa^2 + \frac{2\kappa^2}{9}\left(4-\frac{1}{\eta^2}\right)a^2 + \frac{2\kappa^2}{9}\left(5+\frac{4}{\eta^2}\right)b^2, \notag\\[1mm]
    V_{44}^{(1)} &= \frac{\left(\ell-M+\dfrac{2MK}{3}\right)^2}{r^2} + \frac{2}{3}a^2 - 2\kappa^2 + \frac{2\kappa^2}{9}\left(4+\frac{5}{\eta^2}\right)a^2 + \frac{10\kappa^2}{9}\left(1+\frac{2}{\eta^2}\right)b^2, \notag\\[1mm]
    V_{13}^{(1)} &= V_{23}^{(1)} = \frac{2\sqrt{15}}{3}b', \notag\\
    V_{14}^{(1)} &= \frac{2\sqrt{3}}{3} \left[a' + \frac{M(1-K)}{r}a\right], \notag\\
    V_{24}^{(1)} &= \frac{2\sqrt{3}}{3} \left[a' - \frac{M(1-K)}{r}a\right], \notag\\
    V_{34}^{(1)} &= \frac{2\sqrt{5}}{3} \left(1-\frac{\kappa^2}{\eta^2}\right)ab.
\end{align}

\subsection{\texorpdfstring{\(\rho=-1\)}{rho=-1} block}
In the component order specified by \(u_\ell^{-1}\), the matrix \(V_\ell^{-1}\) is
\begin{align}
    V_\ell^{-1} =
    \begin{pmatrix}
        V_{11}^{(-1)} & 0             & V_{13}^{(-1)} & V_{14}^{(-1)} \\
        0             & V_{22}^{(-1)} & V_{23}^{(-1)} & V_{24}^{(-1)} \\
        V_{13}^{(-1)} & V_{23}^{(-1)} & V_{33}^{(-1)} & V_{34}^{(-1)} \\
        V_{14}^{(-1)} & V_{24}^{(-1)} & V_{34}^{(-1)} & V_{44}^{(-1)}
    \end{pmatrix},
\end{align}
where
\begin{align}
    V_{11}^{(-1)} &= \frac{\left(\ell-1+\dfrac{MK}{3}\right)^2}{r^2} + \frac{2MK'}{3r} + \frac{2}{3}a^2 + \frac{10}{3}b^2, \notag\\[1mm]
    V_{22}^{(-1)} &= \frac{\left(\ell+1+\dfrac{MK}{3}\right)^2}{r^2} - \frac{2MK'}{3r} + \frac{2}{3}a^2 + \frac{10}{3}b^2, \notag\\[1mm]
    V_{33}^{(-1)} &= \frac{\left(\ell+M-\dfrac{2MK}{3}\right)^2}{r^2} + \frac{2}{3}a^2 - 2\kappa^2 + \frac{2\kappa^2}{9}\left(4+\frac{5}{\eta^2}\right)a^2 + \frac{10\kappa^2}{9}\left(1+\frac{2}{\eta^2}\right)b^2, \notag\\[1mm]
    V_{44}^{(-1)} &= \frac{\left(\ell+\dfrac{MK}{3}\right)^2}{r^2} + \frac{10}{3}b^2 - 2\kappa^2 + \frac{2\kappa^2}{9}\left(4-\frac{1}{\eta^2}\right)a^2 + \frac{2\kappa^2}{9}\left(5+\frac{4}{\eta^2}\right)b^2, \notag\\[1mm]
    V_{13}^{(-1)} &= \frac{2\sqrt{3}}{3} \left[a' - \frac{M(1-K)}{r}a\right], \notag\\ V_{23}^{(-1)} &= \frac{2\sqrt{3}}{3} \left[a' + \frac{M(1-K)}{r}a\right], \notag\\
    V_{14}^{(-1)} &= V_{24}^{(-1)} = \frac{2\sqrt{15}}{3}b', \notag\\
    V_{34}^{(-1)} &= \frac{2\sqrt{5}}{3} \left(1-\frac{\kappa^2}{\eta^2}\right)ab.
\end{align}

\section{Bound on the order-three radial profile at \texorpdfstring{\(\kappa=\eta=1\)}{kappa=eta=1}}
\label{app:order-three-BPS-profile-bounds}
In this appendix, we establish the bounds on the radial profiles used in Sec.~\ref{subsec:order-three-analytical-stability}.
We consider regular finite-tension solutions in the order-three vortex family at \(\kappa=\eta=1\).
At \(\eta=1\), the equation for \(b\) decouples, and its regular finite-tension solution is \(b\equiv1\), as shown in Ref.~\cite{Abe:2026zrk}.
The remaining profiles \(K\) and \(a\) therefore satisfy the Bogomol'nyi equations \eqref{eq:order-three-Bogomolny-equations}.
Using the gauge-profile equation \eqref{eq:order-three-profile-equation-K} and the boundary conditions \eqref{eq:order-three-origin-boundary-conditions} and \eqref{eq:order-three-infinity-boundary-conditions}, we show that
\begin{align}
    0 \leq a(r) \leq 1, \qquad r > 0,
    \label{eq:appendix-order-three-BPS-profile-bounds}
\end{align}

Let \(w := K - 1\).
The gauge-profile equation
\eqref{eq:order-three-profile-equation-K} gives
\begin{align}
    \left(\frac{w'}{r}\right)' = \frac{4a^2}{r}w.
\end{align}
If \(w\) were positive on a connected interval, \(w'/r\) would be nondecreasing there. At the left endpoint of such an interval, \(w\) enters the positive region with \(w'\ge0\), so it could neither return to zero nor approach zero at infinity.
This contradicts \(K(\infty)=1\), and hence
\begin{align}
    K(r) \leq 1. \label{eq:appendix-order-three-K-upper-bound}
\end{align}

By uniqueness for the initial-value problem defined by the first equation in \eqref{eq:order-three-Bogomolny-equations}, a zero of \(a\) at any finite \(r>0\) would imply \(a\equiv0\), contradicting \(a(r)\to1\) as \(r\to\infty\).
Hence \(a(r)>0\) for \(r>0\).
Together with \eqref{eq:appendix-order-three-K-upper-bound}, the first Bogomol'nyi equation gives
\begin{align}
    a'(r) = \frac{M(1-K)}{r}a \geq 0.
\end{align}
Thus \(a\) is nondecreasing, and the boundary conditions \(a(r)\to0\) as \(r\to0\) and \(a(r)\to1\) as \(r\to\infty\) give Eq.~\eqref{eq:appendix-order-three-BPS-profile-bounds}.

\section{Explicit radial block operators for the order-two family}
\label{app:order-two-radial-blocks}
We list here the radial block operators defined in Eq.~\eqref{eq:order-two-radial-block-definition}.
Introduce the radial coefficient vectors
\begin{align}
    \widehat{u}_\ell^{\mathrm n} &:= \widehat{\chi}_0,
    \notag\\
    \widehat{u}_\ell^0
    &:=
    \begin{pmatrix}
        \widehat{S}_+^0&
        \widehat{S}_-^0&
        \widehat{\chi}_{+2}&
        \widehat{\chi}_{-2}
    \end{pmatrix}^{\mathsf T},
    \notag\\
    \widehat{u}_\ell^1
    &:=
    \begin{pmatrix}
        \widehat{S}_+^+&
        \widehat{S}_-^+&
        \widehat{\chi}_{+3}&
        \widehat{\chi}_{-1}
    \end{pmatrix}^{\mathsf T},
    \notag\\
    \widehat{u}_\ell^{-1}
    &:=
    \begin{pmatrix}
        \widehat{S}_+^-&
        \widehat{S}_-^-&
        \widehat{\chi}_{+1}&
        \widehat{\chi}_{-3}
    \end{pmatrix}^{\mathsf T}.
\end{align}
The four operators take the form
\begin{align}
    \widehat{L}_\ell^{\mathrm n} &= -\Delta_r + \widehat{V}_\ell^{\mathrm n}(r), &
    \widehat{L}_\ell^\rho &= -\Delta_r\,\mathbf 1_4 + \widehat{V}_\ell^\rho(r), \qquad \rho=0,\pm1, &
    \Delta_r &:= \frac{d^2}{dr^2} + \frac{1}{r}\frac{d}{dr}.
\end{align}
Here \(\widehat{V}_\ell^{\mathrm n}\) is real valued, while \(\widehat{V}_\ell^\rho\), \(\rho=0,\pm1\), are real symmetric matrices.

\subsection{\texorpdfstring{\(\rho=\mathrm n\)}{rho=n} block}
For the \(\rho=\mathrm n\) block,
\begin{align}
    \widehat{V}_\ell^{\mathrm n}
    = \frac{\ell^2}{r^2} - 2\kappa^2
    + 2\kappa^2\left(1+\frac{2}{\eta^2}\right)\widehat{h}^2.
\end{align}

\subsection{\texorpdfstring{\(\rho=0\)}{rho=0} block}
In the component order specified by \(\widehat{u}_\ell^0\), the matrix \(\widehat{V}_\ell^0\) is
\begin{align}
    \widehat{V}_\ell^0 =
    \begin{pmatrix}
        \widehat{V}_{11}^{(0)} & 0            & \widehat{V}_{13}^{(0)} & \widehat{V}_{14}^{(0)} \\
        0            & \widehat{V}_{22}^{(0)} & \widehat{V}_{23}^{(0)} & \widehat{V}_{24}^{(0)} \\
        \widehat{V}_{13}^{(0)} & \widehat{V}_{23}^{(0)} & \widehat{V}_{33}^{(0)} & \widehat{V}_{34}^{(0)} \\
        \widehat{V}_{14}^{(0)} & \widehat{V}_{24}^{(0)} & \widehat{V}_{34}^{(0)} & \widehat{V}_{44}^{(0)}
    \end{pmatrix},
\end{align}
where
\begin{align}
    \widehat{V}_{11}^{(0)} &= \frac{(\ell-1)^2}{r^2} + 4\widehat{h}^2, \notag\\[1mm]
    \widehat{V}_{22}^{(0)} &= \frac{(\ell+1)^2}{r^2} + 4\widehat{h}^2, \notag\\[1mm]
    \widehat{V}_{33}^{(0)} &= \frac{\bigl[\ell+N(1-\widehat{K})\bigr]^2}{r^2} + 2\widehat{h}^2 - 2\kappa^2 + 4\kappa^2\widehat{h}^2, \notag\\[1mm]
    \widehat{V}_{44}^{(0)} &= \frac{\bigl[\ell-N(1-\widehat{K})\bigr]^2}{r^2} + 2\widehat{h}^2 - 2\kappa^2 + 4\kappa^2\widehat{h}^2, \notag\\[1mm]
    \widehat{V}_{13}^{(0)} = \widehat{V}_{24}^{(0)} &= -2\left[\widehat{h}' - \frac{N(1-\widehat{K})}{r}\widehat{h}\right], \notag\\
    \widehat{V}_{14}^{(0)} = \widehat{V}_{23}^{(0)} &= -2\left[\widehat{h}' + \frac{N(1-\widehat{K})}{r}\widehat{h}\right], \notag\\
    \widehat{V}_{34}^{(0)} &= 2(1-\kappa^2)\widehat{h}^2.
\end{align}

\subsection{\texorpdfstring{\(\rho=1\)}{rho=1} block}
In the component order specified by \(\widehat{u}_\ell^1\), the matrix \(\widehat{V}_\ell^1\) is
\begin{align}
    \widehat{V}_\ell^1 =
    \begin{pmatrix}
        \widehat{V}_{11}^{(1)} & 0            & \widehat{V}_{13}^{(1)} & \widehat{V}_{14}^{(1)} \\
        0            & \widehat{V}_{22}^{(1)} & \widehat{V}_{23}^{(1)} & \widehat{V}_{24}^{(1)} \\
        \widehat{V}_{13}^{(1)} & \widehat{V}_{23}^{(1)} & \widehat{V}_{33}^{(1)} & \widehat{V}_{34}^{(1)} \\
        \widehat{V}_{14}^{(1)} & \widehat{V}_{24}^{(1)} & \widehat{V}_{34}^{(1)} & \widehat{V}_{44}^{(1)}
    \end{pmatrix},
\end{align}
where
\begin{align}
    \widehat{V}_{11}^{(1)} &= \frac{\left(\ell-1-\dfrac{N\widehat{K}}{2}\right)^2}{r^2} - \frac{N\widehat{K}'}{r} + 4\widehat{h}^2, \notag\\[1mm]
    \widehat{V}_{22}^{(1)} &= \frac{\left(\ell+1-\dfrac{N\widehat{K}}{2}\right)^2}{r^2} + \frac{N\widehat{K}'}{r} + 4\widehat{h}^2, \notag\\[1mm]
    \widehat{V}_{33}^{(1)} &= \frac{\left[\ell+N\left(1-\dfrac{3\widehat{K}}{2}\right)\right]^2}{r^2} + \frac{3}{2}\widehat{h}^2 - 2\kappa^2 + \frac{\kappa^2}{2}\left(4+\frac{5}{\eta^2}\right)\widehat{h}^2, \notag\\[1mm]
    \widehat{V}_{44}^{(1)} &= \frac{\left[\ell-N\left(1-\dfrac{\widehat{K}}{2}\right)\right]^2}{r^2} + \frac{5}{2}\widehat{h}^2 - 2\kappa^2 + \frac{\kappa^2}{2}\left(4+\frac{3}{\eta^2}\right)\widehat{h}^2, \notag\\[1mm]
    \widehat{V}_{13}^{(1)} &= -\sqrt{3}\left[\widehat{h}' - \frac{N(1-\widehat{K})}{r}\widehat{h}\right], \notag\\
    \widehat{V}_{23}^{(1)} &= -\sqrt{3}\left[\widehat{h}' + \frac{N(1-\widehat{K})}{r}\widehat{h}\right], \notag\\
    \widehat{V}_{14}^{(1)} &= -\sqrt{5}\left[\widehat{h}' + \frac{N(1-\widehat{K})}{r}\widehat{h}\right], \notag\\
    \widehat{V}_{24}^{(1)} &= -\sqrt{5}\left[\widehat{h}' - \frac{N(1-\widehat{K})}{r}\widehat{h}\right], \notag\\
    \widehat{V}_{34}^{(1)} &= \frac{\sqrt{15}}{2}\left(1-\frac{\kappa^2}{\eta^2}\right)\widehat{h}^2.
\end{align}

\subsection{\texorpdfstring{\(\rho=-1\)}{rho=-1} block}
In the component order specified by \(\widehat{u}_\ell^{-1}\), the matrix \(\widehat{V}_\ell^{-1}\) is
\begin{align}
    \widehat{V}_\ell^{-1} =
    \begin{pmatrix}
        \widehat{V}_{11}^{(-1)} & 0             & \widehat{V}_{13}^{(-1)} & \widehat{V}_{14}^{(-1)} \\
        0             & \widehat{V}_{22}^{(-1)} & \widehat{V}_{23}^{(-1)} & \widehat{V}_{24}^{(-1)} \\
        \widehat{V}_{13}^{(-1)} & \widehat{V}_{23}^{(-1)} & \widehat{V}_{33}^{(-1)} & \widehat{V}_{34}^{(-1)} \\
        \widehat{V}_{14}^{(-1)} & \widehat{V}_{24}^{(-1)} & \widehat{V}_{34}^{(-1)} & \widehat{V}_{44}^{(-1)}
    \end{pmatrix},
\end{align}
where
\begin{align}
    \widehat{V}_{11}^{(-1)} &= \frac{\left(\ell-1+\dfrac{N\widehat{K}}{2}\right)^2}{r^2} + \frac{N\widehat{K}'}{r} + 4\widehat{h}^2, \notag\\[1mm]
    \widehat{V}_{22}^{(-1)} &= \frac{\left(\ell+1+\dfrac{N\widehat{K}}{2}\right)^2}{r^2} - \frac{N\widehat{K}'}{r} + 4\widehat{h}^2, \notag\\[1mm]
    \widehat{V}_{33}^{(-1)} &= \frac{\left[\ell+N\left(1-\dfrac{\widehat{K}}{2}\right)\right]^2}{r^2} + \frac{5}{2}\widehat{h}^2 - 2\kappa^2 + \frac{\kappa^2}{2}\left(4+\frac{3}{\eta^2}\right)\widehat{h}^2, \notag\\[1mm]
    \widehat{V}_{44}^{(-1)} &= \frac{\left[\ell-N\left(1-\dfrac{3\widehat{K}}{2}\right)\right]^2}{r^2} + \frac{3}{2}\widehat{h}^2 - 2\kappa^2 + \frac{\kappa^2}{2}\left(4+\frac{5}{\eta^2}\right)\widehat{h}^2, \notag\\[1mm]
    \widehat{V}_{13}^{(-1)} &= -\sqrt{5}\left[\widehat{h}' - \frac{N(1-\widehat{K})}{r}\widehat{h}\right], \notag\\
    \widehat{V}_{23}^{(-1)} &= -\sqrt{5}\left[\widehat{h}' + \frac{N(1-\widehat{K})}{r}\widehat{h}\right], \notag\\
    \widehat{V}_{14}^{(-1)} &= -\sqrt{3}\left[\widehat{h}' + \frac{N(1-\widehat{K})}{r}\widehat{h}\right], \notag\\
    \widehat{V}_{24}^{(-1)} &= -\sqrt{3}\left[\widehat{h}' - \frac{N(1-\widehat{K})}{r}\widehat{h}\right], \notag\\
    \widehat{V}_{34}^{(-1)} &= \frac{\sqrt{15}}{2}\left(1-\frac{\kappa^2}{\eta^2}\right)\widehat{h}^2.
\end{align}

\bibliographystyle{yautphysm}

\bibliography{draft.bib}
\end{document}